# Multiple point adsorption in a heteropolymer gel and the Tanaka approach to imprinting: Experiment and Theory[†]

Kenji Ito[*,1], Jeffrey Chuang[*,2], Carmen Alvarez-Lorenzo[3], Tsuyoshi Watanabe[4], Nozomi Ando[5], Alexander Yu. Grosberg[6]

[1]*National Institute of Advanced Industrial Science and Technology, Tsukuba, Ibaraki 305–8565, Japan,*

[2]*Department of Biochemistry and Biophysics, University of California, San Francisco, California 94143,*

[3]*Departamento de Farmacia y Tecnología Farmacéutica, Facultad de Farmacia, Universidad de Santiago de Compostela, 15706-Santiago de Compostela, Spain,* [4]*Fundamental Research Laboratory, SUMITOMO BAKELITE Co., Ltd., Totsuka-ku, Yokohama-shi, Kanagawa 245–0052, Japan,* [5]*Department of Physics, Cornell University, Ithaca, New York 14853,* [6]*Department of Physics, University of Minnesota, Minneapolis, Minnesota 55455.*

## Abstract

Heteropolymer gels can be engineered to release specific molecules into or absorb molecules from a surrounding solution. This remarkable ability is the basis for developing gel applications in extensive areas such as drug delivery, waste cleanup, and catalysis. Furthermore, gels are a model system for proteins, many of whose properties they can be created to mimic. A key aspect of gels is their volume phase transition, which provides a macroscopic mechanism for effecting microscopic changes. The phase transition allows one to control the gel's affinity for target molecules through tiny changes in the solution temperature, salt concentration, pH, or the like. We summarize recent experiments that systematically characterize the gel affinity as a function of adsorbing monomer concentration, solution salt concentration, and cross-linker concentration, on both sides of the phase transition. We provide a physical theory that explains the results and discuss enhancements via imprinting.

# Contents

## Nomenclature

| | |
|---|---|
| NIPA | $N$-isopropylacrylamide |
| MAPTA$^+$ | methacrylamidopropyl trimethylammonium cation |
| MAPTAC | methacrylamidopropyl trimethylammonium chloride |
| BIS | $N,N'$–methylene–bis(acrylamide) |
| Py-1 | 1–pyrene sulfonate |
| Py-2 | 6,8–dihydroxy-pyrene–1,3–disulfonate |
| Py-3 | 8–methoxy pyrene–1,3,6–trisulfonate |
| Py-4 | 1,3,6,8–pyrene tetrasulfonate |
| $Q$ | The affinity of the gel for the target molecule |
| [Ad] | The concentration of adsorber monomers |
| [Re] | The concentration of replacement molecules, such as salt ions |
| [Xl] | The concentration of cross-linker in the gel |
| $p$ | The number of adsorber monomers binding to the target molecule |
| $\beta$ | The inverse Boltzmann factor $1/k_B T$, in which $k_B$ and $T$ represent the Boltzmann constant and absolute temperature, respectively |
| $\epsilon$ | The binding energy when the target molecule binds a single adsorber monomer, as compared to when it binds a replacement molecule |
| $c$ | A constant dependent on gel composition parameters |
| $S$ | The concentration of binding sites, equivalent to the saturating concentration |
| $K$ | The binding constant for target molecule adsorption by the gel |
| [T$_{ads}$] | The concentration of target molecules adsorbed into the gel |
| [T$_{sol}$] | The concentration of target molecules in the solution surrounding the gel |
| $Z$ | The partition function for target molecule adsorption |
| $Z_p$ | The contribution to the partition function from binding by $p$ adsorber monomers |
| $p_{max}$ | The number of sites on the target molecule to which single adsorber monomers may bind |
| $Q_p$ | The contribution to the affinity from binding by $p$ adsorber monomers |
| $n$ | The number of persistent lengths in an effective chain in the fixed point model |
| $b$ | The persistent length for an effective chain in the fixed point model |
| $a$ | The size of a monomer in an effective chain |
| $m$ | The number of monomers in a persistent length of an effective chain |
| $C_{ad}$ | The molar concentration of adsorber monomers |
| $N_A$ | The Avogadro number |
| $N_{fixed\ points}$ | The total number of fixed points in the gel or the total system volume times the adsorber monomer concentration |
| $\vec{r}$ | The position of a target molecule in the gel |
| $\vec{x_{i_j}}$ | The position of fixed point $i_j$ |
| $\lambda$ | The thermal wavelength of a target molecule |
| $R_p$ | The radius of gyration of a group of $p$ fixed points |
| $\Theta(R_p)$ | The density of states as a function of $R_p$ |
| $\gamma$ | A constant |
| $d$ | The gel diameter in the equilibrium state |
| $d_0$ | $d$ upon polymerization |
| $V$ | The gel volume in the equilibrium state |
| $V_0$ | $V$ upon polymerization |

# 1 Introduction

Thousands of polymeric materials have been invented and tailored to various uses since the discovery of vulcanization by Charles Goodyear in 1839. However impressive, these achievements may one day be dwarfed by the new paradigm – in which polymer molecules act like functioning machines, rather than as building blocks for a material. The use of polymers as so-called 'smart' materials is motivated by the simple observation that biological molecules perform incredibly complex functions. While the goal of engineering polymers as effectively as nature remains somewhat in the realm of science fiction, in recent years many researchers have sought to find or design synthetic polymeric materials capable of mimicking one or another 'smart' property of biopolymers.

Polymer gels are a promising system for such smart functions because of their collapse phase transition, predicted theoretically by Dušek and Patterson in 1968 [1] and then observed experimentally by T. Tanaka in 1978 [2]. Since then, manifestations of this transition have been shown in a variety of circumstances [3–21]. Gel collapse can be driven by any one of the four basic types of intermolecular interactions operational in water solutions and in molecular biological systems [10], namely, by Van der Waals interactions, hydrogen bonds, hydrophobic interactions, and by Coulomb interactions between ionized (dissociated) groups. Gel collapse can be triggered by a variety of external 'stimuli,' including change of temperature (either heating or cooling) [22–27], solvent composition [28–32], pH [33–36], ionic strength [37, 38], and also kinetic influences, such as external electric or magnetic fields [7, 39–42], currents and light [12, 43]. This wealth of properties guarantees rich applications, including super-absorbing materials exemplified by disposable diapers and sanitary napkins [17], materials releasing drugs in a controlled manner [14, 27, 34, 44], catalysis [45], and mimetics of various organic systems [46–56], to name but a few.

A broader perspective of gel collapse is that it is an amplifier allowing conformational changes of macromolecules, huge or tiny as they may be, to be manifested on a macroscopic scale. Once this proposition is accepted, the question arises: what about the intricate conformational changes occurring in biological macromolecules, such as protein folding — can they also be amplified and seen macroscopically in an appropriate gel? In 1992 it was discovered that multiple phases, in addition to the general collapsed and swollen phases, could exist in polymer gels [15]. This work suggested that the phases of a folding protein, and the phenomena of heteropolymer freezing [57–62] in general, could indeed be realized experimentally in a polymer gel. In protein folding, the unique details of the protein's native state, such as its shape and charge distribution, enable it to recognize and interact with specific molecules. Therefore in order to create gels with molecular specificity, it is necessary to design a native state for the gel as well. In recent years, several groups have developed theoretical approaches for designing a heteropolymer's native conformation by choosing its monomer sequence [59, 60]. Such ideas have grown out of research pioneered by the chemist G. Wulff [63–65] to imprint conformations into a polymer sequence. These ideas have recently been applied to realize imprinting of a native state into a heteropolymer gel [66–73].

The imprinting technique in heteropolymer gels can be summarized as follows. A soup of gel monomers (e.g. $N$-isopropylacrylamide, NIPA; an experimental example will be shown in the following section) is placed in solution with charged adsorber monomers (e.g. methacrylamidopropyl trimethylammonium cation, MAPTA$^+$), cross-linker (e.g. $N,N'$–methylene–bis(acrylamide), BIS), and oppositely charged target molecules (e.g. Pyrene Tetra-sulfonate$^{4-}$, Py-4). The target molecules mediate the coulomb interactions between the adsorber monomers. There is an indirect attraction between the adsorber monomers because several of them will tend to cluster around a single multivalent target molecule. Thus the target molecules serve as a sort of glue to stick adsorber monomers together. Polymerization is then initiated while the adsorbers are glued together and the conditions favor the col-lapsed state of the gel. The polymerization fixes the sequence of gel monomers in such a way that the adsorber monomers will cluster together whenever the gel is forced to the collapsed state. After polymerization, the gel can be swollen, to release the target molecules. But whenever the gel is collapsed again, these adsorbing clusters should be recovered, due to the 'imprinting' effect of the target molecules. The recovered clusters will be able to act as molecular recognition sites, specifically attracting molecules with properties similar to those of the gluons.

In this paper we review the first series of experiments related to the imprinting effect in heteropolymer gels sensitive to stimuli. This work focuses on the adsorption properties of random (non-imprinted) heteropolymer gels. Random gels are a large subject in their own right, and they should be understood before imprinting can be placed in the proper context. We first discuss the theoretical ideas behind the adsorption properties of random heteropolymer gels. We then review the experiments that have been done on adsorption of target molecules by random gels, detailing the dependence of their adsorption affinity on adsorber monomer concentration [21], salt concentration [75], and cross-linker concentration [76]. We then compare the properties of random and imprinted gels and show the early successes of the imprinting method [77–84].

The effects observed so far are rather modest and at the moment do not pretend for any comparison with proteins. Nevertheless, it is hoped that imprinting experiments with gels will eventually help capture the principles behind protein self-organization. With this idea in mind, we dedicate this work to the memory of the late Professor Toyoichi Tanaka (1946–2000).

## 2   Theory

In this section, we provide a heuristic derivation of an equation which relates the target molecule affinity of a random gel to several key experimental parameters. This equation will be referred to and discussed in detail in the sections that follow. For convenience, we refer to Eq. (1) as the Tanaka equation since it summarizes a number of findings on gel affinity initiated by T. Tanaka [21, 74–77]. The Tanaka equation for the affinity $Q$ is

$$Q \sim \frac{[\text{Ad}]^p}{p[\text{Re}]^p} \exp(-p\beta\epsilon) \exp\left(-(p-1)c\frac{[\text{Xl}]}{[\text{Ad}]^{2/3}}\right), \qquad (1)$$

where $Q$ is the gel affinity for target molecules. The concentration of adsorbing monomers in the gel is denoted by [Ad], and [Re] is the concentration of replacement molecules, i.e. salt molecules that bind with the target molecule when it is not bound to adsorbing monomers. $p$ is the number of bonds between separate adsorbing monomers and the target molecule. $\beta$ is the Boltzmann factor $1/k_B T$, and $\epsilon$ is the difference in binding energy of a target molecule with (1) an adsorbing monomer or (2) a replacement molecule. [Xl] is the concentration of cross-linker (e.g. BIS). $c$ is a constant which can be estimated from the persistence length and concentration of the main component of the gel chains (e.g. NIPA).

The Tanaka equation can explain how adsorber monomer concentration, salt concentration, and cross-linker concentration control the affinity, as will be described in subsequent experimental sections. Here we present a heuristic explanation of the affinity dependencies for each of these variables and explain what approximations are involved. A more detailed explanation and discussion of the technical issues regarding the Tanaka equation is described in the next section. In this section, we will discuss why adsorption of target molecules is dominated by one value of $p$ at a time, as suggested conceptually by Eq. (1). We will also discuss conditions for the critical switch-like behavior of the gel, in which $p$ changes from 1 in the swollen state to $p_{\max}$ in the collapsed state, where $p_{\max}$ is the number of binding sites on the target molecule.

## 2.1 Langmuir Isotherm and Determination of the Affinity $Q$

Langmuir adsorption [85] has been commonly used to interpret the binding affinity of a gel for a target molecule [86]. The Langmuir isotherm is derived from the concept of a two-state partition function for each binding site, where each binding site is filled or unfilled. There are two terms in the partition function: one proportional to the number of target molecules in solution (analogous to the filled state), and the other independent (analogous to unfilled).

$$[T_{ads}] = S \frac{K[T_{sol}]}{K[T_{sol}] + 1}. \tag{2}$$

Here $[T_{ads}]$ is the concentration of target molecules adsorbed into the gel, $[T_{sol}]$ is the concentration of target molecules in solution, and $S$ is the concentration of binding sites. $K$ is the binding constant, with units of (concentration)$^{-1}$. The overall affinity $Q$ of the binding sites in the gel for the target molecule is defined to be the product of $S$ and $K$, which is dimensionless.

While the Langmuir isotherm is derived from the point of view of the binding sites, the value for $Q$ can be determined from the point of view of a target molecule. Consider the partition function that sums over the different possible states of the target molecule – 0 adsorbers bound, 1 adsorber bound, ... $p_{max}$ adsorbers bound. The partition function will be of the form

$$Z = Z_0 + Z_1 + Z_2 + ... + Z_{p_{max}}, \tag{3}$$

with $Z_p$ indicating the term of the partition function in which a target molecule is bound by $p$ adsorbing monomers. $Z_0$ corresponds to the case of the target molecule being completely unbound.

Looking at the denominator of Eq. 2 for the Langmuir isotherm, we can see that $K[T_{sol}]$ is the term proportional to the fraction of filled binding sites, and 1 is the term proportional to the fraction of vacant binding sites. Therefore $SK[T_{sol}]$ is proportional to the concentration of adsorbed target molecules. Comparing with Eq. (3) we see that $SK[T_{sol}] \propto Z_1 + Z_2 + ... + Z_{p_{max}}$. Meanwhile, the partition function component $Z_0$ for unbound target molecules must be proportional to the number of target molecules in solution, i.e. $[T_{sol}] \propto Z_0$. We can use these two relationships to solve for the affinity $Q = SK$:

$$Q \sim \frac{Z_1 + Z_2 + ... + Z_{p_{max}}}{Z_0}. \tag{4}$$

For convenience, we shall define the quantity $Q_p \equiv Z_p/Z_0$, so that we may consider individual contributions to the affinity in the form $Q = \sum_p Q_p$. We shall next describe a model which allows us to calculate each of the terms $Z_p$.

## 2.2 Fixed-Point Model

We now formulate a model that allows us to calculate the partition function for a single target molecule in solution with a heteropolymer gel. The gel is made up of a primary component (NIPA) as well as some adsorber monomers (MAPTAC). There are target molecules (pyranine) in solution which can diffuse into and out of the gel. The polymer chains in the gel are connected by cross-linker monomers (BIS) into a network.

From a theoretical point of view, the most complex aspect of polymer gels is their network structure. The monomers in the gel can diffuse about to some extent, but are also constrained by the connectivity of the chains in the network. When the cross-linker density is high, there are many constraints on the motion of the monomers. Conversely, at low cross-linker densities, monomers can diffuse more freely. The length scale of monomer localization is determined by the concentration of cross-linker in the gel. We will propose a simplified model to deal with this localization effect. But first, it is important to discuss the motivations for the model we propose.

To form an adsorption binding site of $p$ adsorbing monomers, the monomers have to move in space in order to properly group together. Their motions are severely restricted, because almost every adsorbing monomer in the gel belongs to a sub-chain, which means it is connected by the polymer to two cross-links (there might also be a few adsorbers on the dangling ends, connected to only one cross-link). To understand the amount of freedom afforded to the adsorbing monomers, one has to realize that apart from real cross-links, the freedom of sub-chains is also restricted by the topological constraints, such as entanglements, between polymers. On the other hand, neither real cross-links nor entanglements represent rigid constraints, as they can also move inside the gel.

This situation is far too difficult to address in any systematic theoretical approach. However, from a practical standpoint, what one can do is resort to a more qualitative approximation. We speak about the following mean-field type idea. Let us concentrate on one particular adsorbing monomer in the gel. Owing to a myriad of constraints, it can only access some relatively well defined volume. It is reasonable to assume that in the center of this volume the particle is free to move, but approaching the periphery of its spherical cage the particle feels increasing entropic restrictions. This can be modeled by saying that for each adsorbing monomer in the gel there is a point fixed in space which is the center of the cage, and then there is a free energy potential well (of entropic origin) around this center. We should emphasize that the concept of this center is purely effective – it need not be solidly associated with anything particular in the chemical structure of the gel. This fixed point is as "real" as the self-consistent field itself [92].

To support physical intuition, it is useful to imagine the surrounding potential well by saying that every adsorbing monomer is attached to its corresponding self-consistent center by an effective (also self-consistent) polymer chain, like a dog on a leash (Fig. 1). This polymer chain is not any particular chain in the real gel, but an effective way to describe the entirety of constraints in the gel. It follows from scaling considerations that the length of the effective chain must be of the same order as the distance between cross-linkers in the gel. Thus, it will be a decreasing function of the cross-link density. Clearly, because of the simplifications inherent in this model, it will be useful mainly from a scaling perspective.

Using these motivations, we now describe what we refer to as the "fixed-point model." We imagine that each of the adsorbing monomers is at the end of one of these effective chains. At the other end of the chain is one of these point fixed in space, the positions of which are distributed randomly in the gel. Each chain is assumed to be made of $n$ links, where $n$ is inversely proportional to the density of cross-links in the gel, based on the concept that additional cross-links increase the frustration in the gel. While this assumption ignores dispersity in the location of the adsorber monomers in the actual gel relative to the cross-linking points, it will give the correct scaling dependence on $n$. The parameter $n$ should be proportional to the ratio of main component monomers to cross-linker monomers (e.g. $n \sim$ [NIPA]/[BIS] in a NIPA gel cross-linked with BIS).

An advantage of the fixed-point model is that it allows one to determine the entropic properties of the network using the well-known statistics of polymer chains. Adsorption of target molecules in the gel will deform the chain network, and the accompanying entropy loss will be analyzed via the entropy of gaussian chains. This entropic effect will be affected not only by the cross-linker density, but also by the density of adsorber monomers – which are implicit in the definition of a fixed point. Note that this density can be adjusted via the gel swelling phase transition. In the swollen state there will be a low density of fixed points, and in the collapsed state there will be a high density.

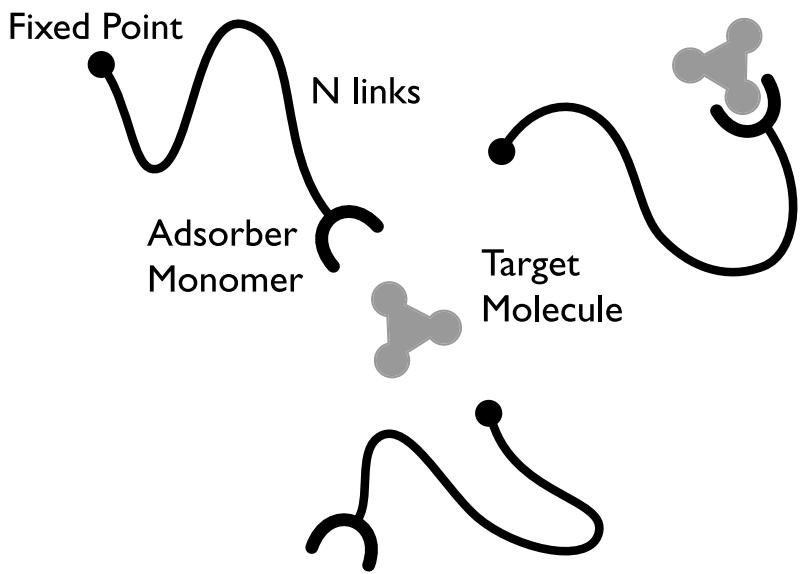

Figure 1: Schematic drawing for the model of a gel with fixed points for cross-linking constraints. The fixed-point model replaces these constraints by placing each adsorber monomer at the end of a finite chain of length $n$, the value of which is inversely proportional to the cross-linker density.

## 2.3   Heuristic Derivation and Implications of the Tanaka Equation

We now give a qualitative explanation of the various terms in the Tanaka equation, which describes the affinity of a gel for a target molecule as it is bound by $p$ adsorbing monomers. Experiments which have assessed the Tanaka equation will be described in subsequent sections.

We first explain the power-law dependence of the affinity $Q$ on [Ad]. For a target molecule to be adsorbed, the $p$ adsorbing monomers must be clustered together. The probability of such a cluster existing at a given point in a random gel is a product of the probabilities for each of the adsorbing monomers. Therefore the dependence goes as $[Ad]^p$. Each of these clusters requires $p$ adsorbing monomers, hence $Q$ is proportional to $1/p$.

The affinity is proportional to $[Re]^{-p}$ because these replacement molecules (typically salt ions) act as competitors to the adsorbing monomers. In solution, a target molecule may either be adsorbed into the gel or bound by $p$ replacement molecules. Binding to the replacement molecules prevents adsorption by the gel. In order for a target to be bound to $p$ replacement molecules, the replacements must cluster around the target. This creates a power-law dependence similar to that for target molecule adsorption by adsorbing monomers, but with an opposite sign exponent.

The energetic attraction of a target molecule to $p$ adsorbing monomers is encapsulated in the term $\exp(-p\beta\epsilon)$. This is a Boltzmann probability based on a binding energy $\epsilon$ per adsorbing monomer.

The dependence of the affinity, $Q$, on the cross-links can be explained as follows. The adsorber units in the gel can move rather freely within a certain volume determined by the cross-link density. Below a certain length scale associated with the cross-link density, the gel behaves like a liquid, allowing the adsorber groups to diffuse virtually freely. Beyond that length scale, however, the gel behaves as an elastic solid body. The adsorber units cannot diffuse further than that length scale. As shown earlier in Fig. 1, we assume that each adsorber is at one end of a fictitious Gaussian chain with a length half the average polymer length between the nearest cross-links (Xl):

$$l = nb = ([\text{NIPA}]/2[\text{Xl}])\,a \tag{5}$$

Here $n$ is the number of monomer segments of persistent length $b$ contained in the chain. If there are $[\text{NIPA}]/2[\text{Xl}]$ monomers between the cross-link and an adsorbing monomer group, then in Eq. (5), $n = [\text{NIPA}]/2m[\text{Xl}]$ and $b = ma$, where $m$ is the number of monomers involved in the persistent length and $a$ is the length of each monomer. At a concentration of $[\text{Ad}]$ of adsorbing monomers, the average spatial distance between adsorbing monomers is $R = [\text{Ad}]^{-1/3}$. (For a molar concentration $C_{\text{ad}}$ this corresponds to $R = 1$ cm$/(C_{\text{ad}}N_{\text{A}})^{1/3}$, where $N_{\text{A}}$ is the Avogadro number.) This fictitious Gaussian chain represents the restricted ability of the adsorber groups to diffuse within a certain volume in the gel. We expect that the probability for two adsorber monomers to meet should be proportional to the Boltzmann factor of the entropy loss associated with the formation of one pair of adsorbers.

$$P = P_0 \exp(-R^2/nb^2) = P_0 \exp\left(-c[\text{Xl}]/[\text{Ad}]^{2/3}\right), \tag{6}$$

where the quantity $c$ is determined by the persistence length, the number of monomers in a persistence length, and the concentration of the main component of the chains through the relation

$$c = \frac{2m}{[\text{NIPA}]b^2}. \tag{7}$$

Since the adsorption of a divalent target by two adsorbers brings together each end from two fictitious Gaussian polymers, the affinity should be proportional to this probability. If more than two contact points are expected, the equation can be generalized as

$$Q \propto \exp\left(-(p-1)c[\text{Xl}]/[\text{Ad}]^{2/3}\right). \tag{8}$$

If the target molecule is adsorbed only by a single contact ($p = 1$), $Q$ should be independent of the cross-linker concentration. However, if the target binding site requires several ($p > 1$) adsorber monomers, the cross-links will frustrate the formation of the binding site, and the frustration will increase with $p$.

Since the Tanaka equation is an expression for the total affinity of a gel for the target molecule, it implies that adsorption is dominated by a single value of $p$. This can be understood by considering whether the attraction of the target molecule to adsorber monomers (due to energetic and concentration effects) is stronger than the repulsion due to the entropy loss required to deform the gel. If attraction is sufficiently favored, then the target will be bound by as many adsorber monomers as possible ($p = p_{\text{max}}$). However, if the entropy loss to deform the gel is stronger, then the only adsorption will be by single adsorber monomers, which would not require deformation of the network ($p = 1$).

The basic concept of gels as smart materials is that they will have high affinity for the target in the collapsed state, but low affinity in the swollen state. By controlling the phase transition of the gel, one will be able to create a switch-like behavior in the affinity. The Tanaka equation allows us to predict the composition of gels which will drastically change affinity during the gel phase transition. In order for the gel to have a low affinity in the swollen state, the adsorber monomers should have only a weak attraction to the target molecules, i.e. any adsorption should be single-handed ($p = 1$). To have a high gel affinity, adsorption in the collapsed phase should involve as many adsorber monomers as possible ($p = p_{\text{max}}$). To effect a change in $p$, one must change the relative strengths of terms with different $p$. The $p$ value transition should occur where the entropic and energetic contributions to the affinity are equal, i.e. the crossover should occur when

$$\ln([\text{Ad}]/[\text{Re}]) \approx ([\text{Ad}]^{2/3}nb^2)^{-1} + \beta\epsilon. \tag{9}$$

Thus the Tanaka equation provides us with a condition that allows us to design a gel with switch-like behavior in its affinity. At the gel volume phase transition, the concentrations of adsorber, cross-linker, and salt, as well as the temperature of the system may be altered to control this switch. To make a gel in which $p$ changes across the volume phase transition, one should choose components so that Eq. (9) is valid at the center of the transition. The experiments discussed in the following sections use gels in which $p$ changes across the phase transition. However, designing such gels still requires a significant amount of testing, since it is difficult to know the exact value for the binding energy $\epsilon$ a priori.

## 2.4 Discussion of the Tanaka Equation

While most of the ideas behind the Tanaka equation can be understood by the heuristic arguments of the preceding section, it has been difficult to completely derive the Tanaka equation formally. There are four issues regarding its applicability that require some discussion. They are:

1. Delta function approximation for the distribution of fixed points.

2. Adsorber monomer dependence of the saturation level $S$.

3. Domination of binding by a single value of $p$.

4. Modifications to $n$ by effective cross-linking.

In order to elucidate these issues, we write out a formal expression for the partition function $Z$. The partition function can be written as a sum of contributions from terms corresponding to the number of adsorbing monomers $p$ binding the target molecule:

$$Z = \sum_{p=0}^{p_{\max}} Z_p. \tag{10}$$

The terms $Z_p$ have contributions from the binding energy of an adsorber to the target molecule, the displacement entropy of the replacement molecules, the binding entropy of the adsorber monomers, and the deformation entropy of the polymer network.

If one follows the formulation of the partition function of an ideal gas [87], there is also a trivial contribution due to different possible momenta of the target molecule: $Z_p \propto \frac{1}{\lambda^3}$, where $\lambda$ is the thermal wavelength of the gas of target molecules. The value of $\lambda$ is irrelevant to the affinity, since $Q$ is a sum of terms $Z_p/Z_0$ which have no $\lambda$ dependence. This $\lambda$ contribution is only important in order to allow for the partition function to be dimensionless (e.g. $Z_0 = V/\lambda^3$, where $V$ is the volume of the system).

There is a binding energy associated with the adsorption of the adsorber monomer to the target molecule, which we define to be $\epsilon$. $\epsilon$ is the difference in binding energy of an adsorbing monomer to the target molecule and a replacement molecule to the target molecule. This definition is necessary because a replacement molecule (i.e. a salt ion) must first be displaced from the target in order for an adsorber monomer to be bound.

$$Z_p \propto \exp(-p\beta\epsilon), \tag{11}$$

where we have assumed that each of the binding sites gives an identical contribution $\epsilon$ to the energy. In a system with heterogeneous binding sites or cooperative effects, multiple values of $\epsilon$ could be used to generalize.

As mentioned in the last section, the affinity is proportional to $[\text{Re}]^{-p}$ because replacement molecules act as competitors to the adsorbing monomers. Binding of a replacement molecule to a gel reduces the entropy of the replacement molecule, which decreases the partition function by a factor proportional to the concentration of replacement molecules [Re]. Therefore the term $Z_p$ has the [Re] dependence

$$Z_p \propto [\text{Re}]^{-p}. \tag{12}$$

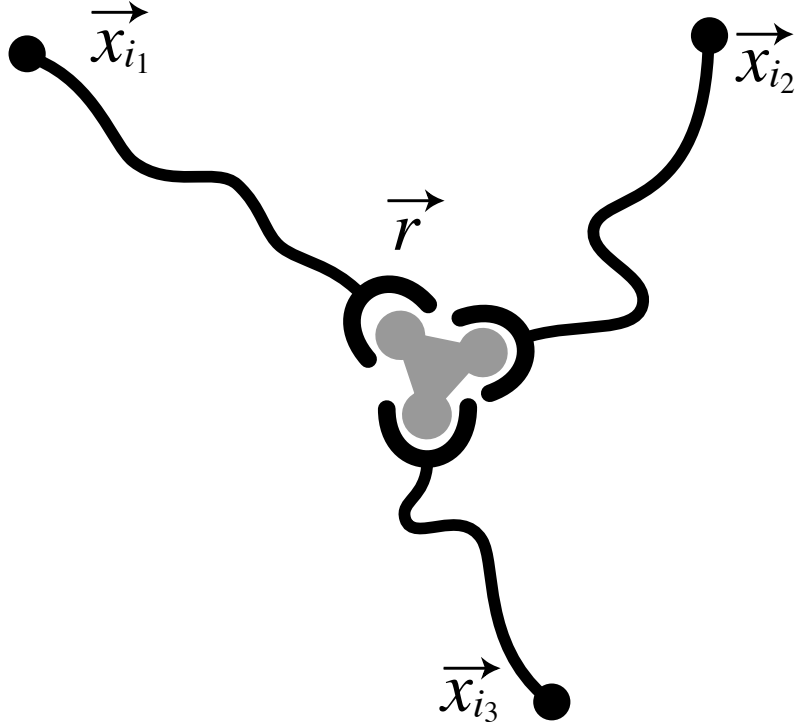

Figure 2: Capture of a target molecule by three adsorbing monomers. $\vec{x_{i_1}}$, $\vec{x_{i_2}}$, and $\vec{x_{i_3}}$ refer to the positions of the respective fixed points. The target molecule is located at $\vec{r}$. Each adsorbing monomer can bind to the target molecule with an energy $\epsilon$.

Finally, the partition function contribution $Z_p$ is affected by the binding of the adsorber monomers to the target molecules, which causes stretching of the chains which connect the adsorber monomers to the fixed points (Fig. 2).

$$Z_p \propto \sum_{i_1 < i_2 < \ldots < i_p}^{N_{\text{fixed points}}} \int d\vec{r} \prod_{j=1}^{p} \frac{\exp\left(-(\vec{x_{i_j}} - \vec{r})^2 / nb^2\right)}{(\pi nb^2)^{3/2}} \tag{13}$$

This equation requires some explanation. Here $\vec{r}$ is the position of the target molecule. The target is bound by $p$ chains, which are indexed by the symbol $j$. Each of these chains have an adsorber monomer glued to the target at one end and have a fixed point at the other. There are many possible adsorber monomers which may attach to a target molecule, and each of these adsorber monomers has a one-to-one association with a fixed point. To account for the large number of fixed points in the gel we sum over all possible ways in which $p$ fixed points can be chosen from the set of all fixed points. Let the fixed points attached to chains $1, 2, ..., p$ be indexed by $i_1, i_2, ...i_p$, respectively. Then we define $\vec{x_{i_j}}$ to be the location of the fixed point indexed by $i_j$. The total number of fixed points in the gel $N_{\text{fixed points}}$ is simply the total system volume times the adsorber monomer concentration.

When the target molecule is bound to the gel, the chain stretching is equal to the vector connecting the target molecule and its fixed point. Thus the stretching of chain $j$ is equal to the vector $(\vec{x_j} - \vec{r})$. We can account for the entropic penalty when chains are stretched by using a standard Gaussian formalism for chain entropy [92]. If a chain has extension $\vec{y}$, then the associated entropic factor is $\frac{\exp(-y^2/nb^2)}{(\pi nb^2)^{3/2}}$, where we have assumed each of these chains is made of $n$ links of persistent length $b$. The product shown in Eq. (13) is the result when all $p$ chains are considered. We have followed the form for the chain entropy used in Ref. [79] for consistency with previous work. Technically speaking, the entropic factor should be of the form $\frac{\exp(-3y^2/2nb^2)}{(2\pi nb^2/3)^{3/2}}$ to agree with Ref. [92], however this discrepancy is relatively unimportant as it only alters our effective value of $n$ by the relation $n \to 2n/3$. As we will see in the following section, other approximations are likely to affect the chain entropy term at least as significantly.

Combining these contributions to $Z_p$ we arrive at the following equation, which describes all of the phenomena necessary for understanding the Tanaka equation.

$$Z_p = (1/\lambda^3)\frac{\exp(-p\beta\epsilon)}{[\text{Re}]^p} \sum_{i_1 < i_2 < ... < i_p}^{N_{\text{fixed points}}} \int d\vec{r} \prod_{j=1}^{p} \frac{\exp\left(-(\vec{x_{i_j}} - \vec{r})^2/nb^2\right)}{(\pi nb^2)^{3/2}} \tag{14}$$

The contribution of $Z_p$ to the affinity is simply $Q_p = Z_p/Z_0$.

To simplify the expression for $Z_p$ we can integrate over the different possible locations $\vec{r}$ of the target molecule. This can be done by converting all of the chain entropy contributions into a single exponential quadratic in $\vec{r}$ and then performing a Gaussian integration. Since we have already handled the trivial case $p = 0$, we will also assume below that $p \geq 1$.

$$
\begin{aligned}
\prod_{j=1}^{p} \exp\left(-(\vec{x_{i_j}} - \vec{r})^2/nb^2\right) &= \exp\left(-\frac{1}{nb^2}\sum_{j=1}^{p}(\vec{x_{i_j}} - \vec{r})^2\right) \\
&= \exp\left(-\frac{1}{nb^2}\sum_{j=1}^{p}(\vec{x_{i_j}}^2 - 2\vec{x_{i_j}} \bullet \vec{r} + \vec{r}^2)\right) \\
&= \exp\left(-\frac{p}{nb^2}\left[\frac{\sum_{j=1}^{p}\vec{x_{i_j}}^2}{p} - 2\frac{\sum_{j=1}^{p}\vec{x_{i_j}} \bullet \vec{r}}{p} + \vec{r}^2\right]\right) \\
&= \exp\left(-\frac{p}{nb^2}\left[\frac{\sum_{j=1}^{p}\vec{x_{i_j}}^2}{p} - \left(\frac{\sum_{j=1}^{p}\vec{x_{i_j}}}{p}\right)^2 + \left(\vec{r} - \frac{\sum_{j=1}^{p}\vec{x_{i_j}}}{p}\right)^2\right]\right)
\end{aligned}
\tag{15}
$$

We now use the standard Gaussian integration formula $\int_{-\infty}^{\infty} \exp(-\alpha y^2)d\vec{y} = (\pi/\alpha)^{3/2}$ with the substitution $\vec{y} = \vec{r} - \frac{\sum_{j=1}^{p}\vec{x_{i_j}}}{p}$, to perform the integration on $\vec{r}$, arriving at

$$\int d\vec{r} \prod_{j=1}^{p} \exp\left(-(\vec{x_{i_j}} - \vec{r})^2/nb^2\right) = (\frac{\pi nb^2}{p})^{3/2} \exp\left(-\frac{p}{nb^2}\left[\frac{p\sum_{j=1}^{p}\vec{x_{i_j}}^2 - (\sum_{j=1}^{p}\vec{x_{i_j}})^2}{p^2}\right]\right). \tag{16}$$

One way to simplify Eq. (16) is to note that the exponent is closely related to the radius of gyration of the fixed points at the $\vec{x_{i_j}}$. The radius of gyration $R_p$ of a collection of $p$ points at positions $\vec{r_1}, \vec{r_2}, ..., \vec{r_p}$ is defined [92] to be

$$R_p \equiv \sqrt{\frac{\sum_{1 \leq j < k \leq p} (\vec{r_j} - \vec{r_k})^2}{p^2}}.$$ (17)

If we square this expression, and rearrange the terms, we have

$$R_p^2 = \frac{p \sum_{j=1}^{p} \vec{r_j}^2 - (\sum_{j=1}^{p} \vec{r_j})^2}{p^2}.$$ (18)

This expression for the radius of gyration can be substituted into the exponent of Eq. (16). Using $\mathcal{Z}_0 = V/\lambda^3$, we arrive at the affinity

$$Q_p = \frac{\exp(-p\beta\epsilon)}{V[\text{Re}]^p} (\pi n b^2)^{-3p/2} (\frac{\pi n b^2}{p})^{3/2} \sum_{i_1 < i_2 < ... < i_p}^{N_{\text{fixed points}}} \exp(-pR_p^2/nb^2).$$ (19)

The gel entropy loss due to adsorption has a simple physical interpretation. It is a sum of gaussian weights from all the possible combinations of fixed points, where the exponent is determined by the radius of gyration of each set of fixed points. Note that for $p = 1$ there is no chain entropy penalty, since the radius of gyration of a single point is by definition equal to zero. With this complete physical interpretation in hand, we now proceed to discuss four important approximations that relate Eq. (19) to the heuristically derived Tanaka equation, Eq. (1).

### 2.4.1 Delta function approximation for the distribution of fixed points.

Eq. (19) allows us to interpret the effects of gel deformation on the entropy. However, proceeding from this expression to one containing experimental parameters is still a difficult task. Although the fixed points are uniformly distributed throughout the gel, understanding the entropy loss requires one to account for the correlations in positions between fixed points. Adsorption of a target molecule involves several fixed points interacting with each other via their attached adsorbing monomers; therefore these correlations are essential for calculating the entropy. The Tanaka equation deals with these correlations by making a rather large approximation which will be described below.

The radius of gyration $R_p$ contains the necessary correlation information for characterizing the gel entropy, therefore for conceptual simplicity, we express the affinity $Q_p$ in terms of an integration over all possible radii of gyration:

$$Q_p = \frac{\exp(-p\beta\epsilon)}{V[\text{Re}]^p} (\pi n b^2)^{-3p/2} (\pi n b^2/p)^{3/2} \int dR_p \Theta(R_p) \exp(-pR_p^2/nb^2),$$ (20)

where the notation $\Theta(R_p)$ is the density of states as a function of the radius of gyration $R_p$ of the $p$ fixed points. The distribution function $\Theta(R_p)$ in Eq. (20) uses the location information of all the fixed points in the gel. $\Theta(R_p)$ therefore contains the affinity dependence on the adsorber monomer concentration. Changes in affinity due to the volume phase transition will manifest themselves through this term, since volume changes will alter the adsorber monomer concentration. At higher concentrations of adsorbing monomer the distribution will be weighted toward small values of $R_p$. Conversely, at low concentrations of [Ad] the distribution will be weighted toward large values of $R_p$.

The Tanaka equation deals with the distribution function $\Theta(R_p)$ by assuming that $\Theta(R_p)$ has a delta-function like dependence at the value $R_p \sim [\text{Ad}]^{-1/3}$ [77, 88]. The explanation for this dependence on [Ad] is that chains bound to the same target molecule will be stretched by the typical distance that separates two nearby fixed points in space, which is $[\mathrm{Ad}]^{-1/3}$. This is a mean-field approximation which neglects the variations in the distribution function, which naturally can range over any $R_p$ from 0 to the size of the gel. The Tanaka equation approximation is based on the idea that there are many sets of adsorber monomers that work together locally to form binding sites for the target molecules, since the probability for chain stretching drops off exponentially with distance. Therefore binding sites will tend to be made up from adsorber monomers whose fixed points are close to each other. There should be an averaging of the behavior over these local domains, making contributions to the distribution function for $R_p > [\mathrm{Ad}]^{-1/3}$ negligible.

This delta-function approximation allows one to compute the change in gel entropy due to adsorption of a target molecule, though its approximate nature suggests that the Tanaka equation is most useful for a scaling analysis. The form of $\Theta(R_p)$ implied by the Tanaka equation is

$$\Theta(R_p) = \delta(R_p - [\mathrm{Ad}]^{-1/3})[\mathrm{Ad}]V\left([\mathrm{Ad}]n^{3/2}b^3\right)^{p-1}. \tag{21}$$

The delta function indicates that grouping of adsorbing monomers occurs on a scale $[\mathrm{Ad}]^{-1/3}$. $[\mathrm{Ad}]V$ is the number of potential adsorbing monomers to which the target molecule can form a single contact. $([\mathrm{Ad}]n^{3/2}b^3)^{p-1}$ is the probability – given a target molecule adsorbed by an adsorber monomer – that $(p-1)$ more adsorbing monomers are close enough to bind the target as well. Since the chain stretching length is of order $n^{1/2}b$, other adsorbing monomers are "close enough" if they are within a volume $n^{3/2}b^3$ near the target molecule. One special exception to this location probability $([\mathrm{Ad}]n^{3/2}b^3)^{p-1}$ is in collapsed imprinted gels, in which fixed points are clustered, meaning $\Theta(R_p) \propto [\mathrm{Ad}]$ instead. That phenomenon will be described further in the section, which discusses gel imprinting.

We can insert the random gel expression for $\Theta(R_p)$ into Eq. (20), to arrive at the affinity. Simultaneously we make two other modifications. First, as mentioned earlier, when $p = 1$, there should be no chain entropy term. This is because at fixed $R_p$, only $p-1$ of the $p$ chains are independent. So the chain entropy behaves as though only $p-1$ chains are stretched. To show this more explicitly, we multiply the affinity by a constant factor which effectively shifts $p \to (p-1)$. Second we make the substitution $c[\mathrm{Xl}] = 1/nb^3$, as described in the heuristic derivation of the Tanaka equation. This results in

$$Q_p \sim \frac{\exp(-p\beta\epsilon)}{[\mathrm{Re}]^p}[\mathrm{Ad}]^p(\pi/p)^{3/2}\exp\left(-(p-1)c\frac{[\mathrm{Xl}]}{[\mathrm{Ad}]^{2/3}}\right). \tag{22}$$

This simplified version of $Q_p$ has all the key dependences of the Tanaka equation. We previously explained the binding energy and replacement molecule dependence. Here we have shown that the explicit adsorber monomer dependence is $[\mathrm{Ad}]^p$ and the chain entropy dependence is $\exp(-(p-1)c\frac{[\mathrm{Xl}]}{[\mathrm{Ad}]^{2/3}})$. However, since we used the significant approximation that the fixed points are distributed at a characteristic radius of gyration from each other $[\mathrm{Ad}]^{-1/3}$, the Tanaka equation should primarily be considered as a scaling relation, which may be off by small numerical factors.

### 2.4.2 Adsorber Monomer Dependence of the Saturation Level $S$

The binding energy, replacement molecule, adsorber monomer, and cross-linker dependence as given in the Tanaka equation Eq. (1) are all explained by Eq. (22), but the saturation level $S$ is not. According to the heuristic explanation for the Tanaka equation, one would expect $Q_p \propto 1/p$. In order to explain this effect, we discuss the delta function approximation of the previous section.

In actuality, Eq. (22) is a description of the affinity only at low concentration of adsorber monomer. The justification for the $[Ad]^p$ dependence in Eq. (21) was based on estimating the number of (scarce) multipoint binding sites – which suggests that the saturation value $S$ should scale as $[Ad]^p$. However, if all adsorber monomers participate in multipoint binding sites, the saturation value must be $[Ad]/p$. This inconsistency stems from a change in the number of effective binding sites as the adsorber monomer concentration is increased. At low concentrations of $[Ad]$, only a small fraction of the adsorber monomers will be close enough together to participate in multipoint binding sites, which is the situation described by the delta-function approximation of Eq. (21). At high adsorber concentrations all of the monomers will be involved in multipoint binding, corresponding to the saturation value $S \propto [Ad]/p$.

The Tanaka equation combines these two situations into a single equation. This allows one to use the low $[Ad]$ regime to observe the multipoint adsorber binding, but to also obtain the correct saturation value at high $[Ad]$. The overall dependence $Q \propto \frac{[Ad]^p}{p}$ given in the Tanaka equation is therefore a phenomenological joining of two separate situations. The reason that such approximations are necessary is that characterizing the $S$ vs. $[Ad]$ curve at arbitrary $[Ad]$ is a complex problem. Fully characterizing $S$ would involve enumerating the numbers of $p$-handed binding sites over random fixed point distributions, a very difficult problem in clustering theory.

### 2.4.3 Domination of binding by a single value of $p$

The Tanaka equation suggests that the affinity can be understood by considering only a single value of $p$ at a given time. However, in reality, there must be a variety of different types of adsorption occurring simultaneously, with multiple values of $p$. The balance of types of adsorption will affect the observed saturation value $S$ measured in Langmuir adsorption.

The single $p$ behavior implied by the Tanaka equation can be justified by considering the relative contributions of each value of $p$ to the affinity. Taking the logarithm of the individual terms in the affinity $Q_p$, we have

$$\ln(Q_p) = p \left[ \ln \frac{[Ad]}{[Re]} - \beta\epsilon - c\frac{[Xl]}{[Ad]^{2/3}} \right] + c\frac{[Xl]}{[Ad]^{2/3}} - \ln p. \tag{23}$$

If one ignores the $\ln p$ term, which has only a minor contribution, then this equation has the generic form

$$\ln(Q_p) = pA + B, \tag{24}$$

where $A = \ln \frac{[Ad]}{[Re]} - \beta\epsilon - c\frac{[Xl]}{[Ad]^{2/3}}$ and $B = c\frac{[Xl]}{[Ad]^{2/3}}$. $Q_p$ will be maximized at either $p = 1$ or $p = p_{\max}$ since it is monotonic in $p$. For $A > 0$, we will have $p = p_{\max}$ and for $A < 0$, we will have $p = 1$. There is an exponential dependence of $Q_p$ on the value of $p$, meaning that if $A$ has a large magnitude, then the adsorption should be dominated by the single value $p = 1$ or $p = p_{\max}$. This is the substance of the single $p$ approximation implicit in the Tanaka equation – that $|A| > 1$.

This analysis predicts the experimental conditions under which binding will be single ($p = 1$) or multi-handed ($p = p_{\max}$). To make a gel with $p = 1$ in the swollen state and $p = p_{\max}$ in the collapsed state, $A$ must change sign during the volume phase transition. However, it is difficult to construct such a gel from first principles, since accurate values of all the parameters are not easily obtainable.

### 2.4.4 Modifications to $n$ by effective cross-linking

There are two issues which may modify the value of $n$. If the gel has adsorbed target molecules, then the adsorption complexes act as cross-links as well, which would modify the functional form of $n$ such that $n \propto$ $(\gamma[\mathrm{Ad}] + [\mathrm{Xl}])^{-1}$, where $\gamma$ is a constant. In the experiments described in later sections, the concentration of cross-linker is much higher than the concentration of adsorbing monomer (e.g. $[\mathrm{BIS}] \gg [\mathrm{MAPTAC}]$); so the explicit cross-linker dependence will dominate. However, it is conceivable that for a gel with low enough concentration of cross-linker, target molecule adsorption may be more important. Also, the Tanaka equation ignores topological effects, which may create effective cross-links due to knots formed between chains.

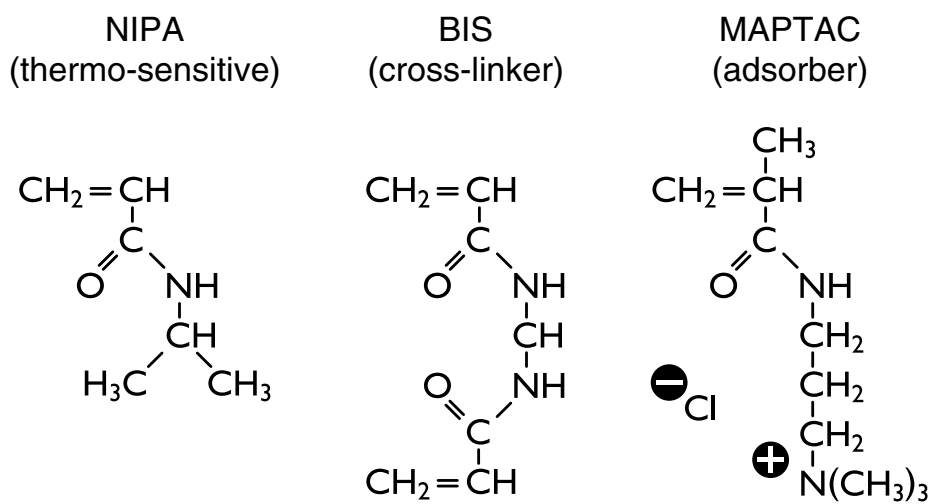

Figure 3: Chemical structure of monomers.

# 3 Experimental Assessment of the Tanaka Equation

Having discussed the concepts underlying the affinity of a gel for target molecules, we now describe several experiments which assess the functional form of the Tanaka equation. Experiments have been performed which test the affinity dependence on adsorber monomer concentration, salt (replacement molecule) concentration, and cross-linker dependence. At the head of each section, we have reiterated the appropriate functional form of the dependence predicted by the Tanaka equation.

## 3.1 Methods

### 3.1.1 Gel preparation

The gels were prepared by free radical polymerization [89] using 6 M $N$-isopropylacrylamide (NIPA, which was kindly supplied from Kohjin Co., Japan), 0–120 mM methacrylamidopropyl trimethylammonium chloride (MAPTAC, Mitsubishi Rayon Co. Ltd., Japan) and 5–200 mM cross-linker $N,N'$–methylene–bis(acrylamide) (BIS, Polysciences Inc., PA) as shown in Fig. 3. After the monomers were dissolved in dimethyl-sulfoxide, 10 mM 2,2′–azobisisobutyronitrile (AIBN, initiator) was added, and the solutions were immediately transferred to test tubes in which micropipettes of inner diameter of approximately 0.5 mm were placed. The solutions filled the micropipettes, and were then degassed under vacuum for a few seconds. The polymerization was carried out at 60 °C for 24 hours. After gelation was completed, the micropipettes were crushed. To remove unreacted chemicals, the gels were washed with large amounts of 100 mM hydrochloric acid (HCl) and sodium chloride (NaOH) aqueous solutions and then with deionized distilled water for ten days. Finally, the gels were removed from the solution and dried under vacuum for one week.

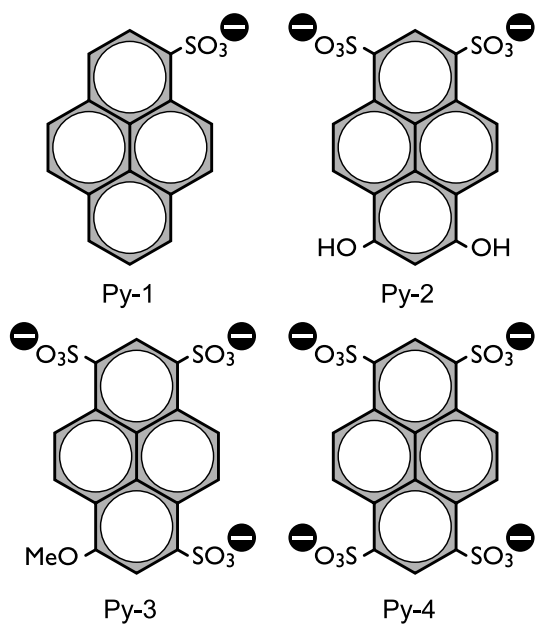

Figure 4: Chemical structure of targets.

### 3.1.2 Degree of swelling

Equilibrium diameters $d$ of the cylindrical gels in water were measured using a microscope equipped with a CCD video camera. The degree of swelling $V/V_0$ was expressed as:

$$\frac{V}{V_0} = \left(\frac{d}{d_0}\right)^3,\tag{25}$$

where $d_0$ was the gel diameter upon polymerization. It was confirmed that the swelling ratio for all collapsed gels at 60 °C was 1, meaning the same as that at preparation, irrespective of the experimental conditions such as adsorber concentration (0–120 mM), salt concentration (27–200 mM), and cross-linker concentration (5–200 mM).

### 3.1.3 Adsorption Studies

As adsorbates or target molecules, several different types of pyrene sulfonate derivatives were used: 1–pyrene sulfonic acid sodium salt (Py-1·Na), 6,8–dihydroxy-pyrene–1,3–disulfonic acid disodium salt (Py-2·2Na), 8–methoxy pyrene–1,3,6–trisulfonic acid trisodium salt(Py-3·3Na), and 1,3,6,8–pyrene tetrasulfonic acid tetrasodium salt (Py-4·4Na), portraying in Fig. 4. These chemicals present 1 (Py-1), 2 (Py-2), 3 (Py-3) or 4 (Py-4) anionic charges, which can interact electrostatically with a cationic charged site such as on MAPTA⁺.

Pieces of cylindrical gel (5–20 mg dry weight) were placed in 2 or 4 mL target aqueous solution (the concentration of the targets ranged from 2 $\mu$M to 0.5 mM). The solutions also contained NaCl of a prescribed concentration to provide monovalent chloride ions to replace the target molecules, namely, the Donnan potential, which could have immobilized the charged target molecules at the interface between the gel and the outer solution, is insignificant for the adsorption process. The samples were kept swollen (20°C) or shrunken (60 °C) for 48 hours. Equilibrium concentration of the target molecules in the medium was measured spectrophotometrically. The amount of target adsorbed by the samples was then evaluated as the difference between the initial and the final quantities in the medium. These amounts were then converted to units of moles of target molecule per volume of gel at the time of synthesis. The adsorption isotherms were analyzed in terms of the Langmuir equation, Eq. (2). Once again, in this equation $[T_{ads}]$ is the amount of target adsorbed per unit volume of gel in the shrunken state, $[T_{sol}]$ is the final equilibrium concentration in the solvent, $S$ is the number of adsorbing sites per unit volume of gel, and $K$ is the affinity of one adsorption site for a target molecule. From the slope and the intercept at zero $[T_{sol}]$ we can deduce both $S$ and $K$, and the overall affinity $Q$. Typical adsorption results are shown in Fig. 5, which displays the adsorber dependence of the adsorption of Py-3 for a collapsed gel. All the adsorption isotherms were well fit by Eq. (2).

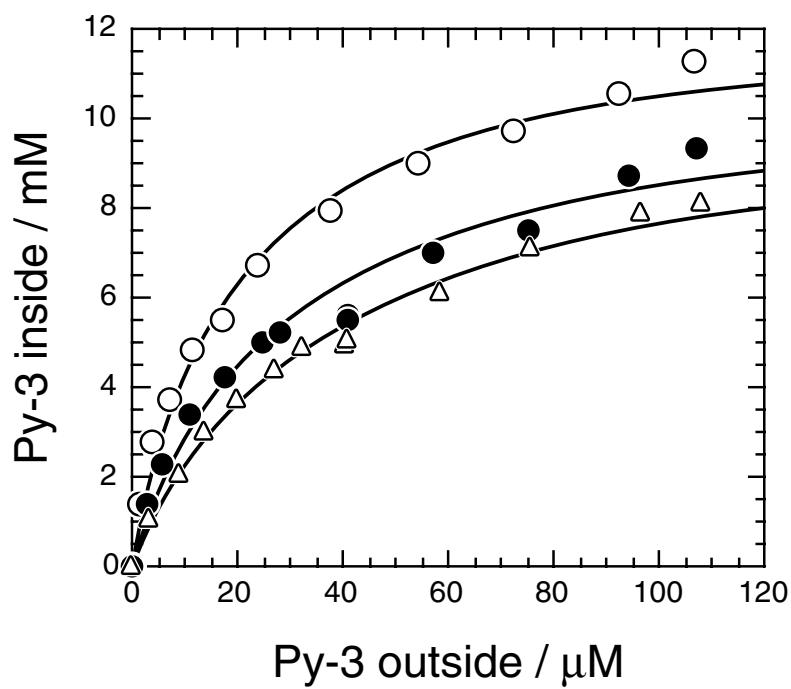

Figure 5: Typical adsorption isotherms of the collapsed gel at 60 °C for the adopted binding systems. The experimental conditions are as follows: The target molecule is Py-3. The concentrations of the main monomer (NIPA), the adsorber (MAPTAC) and the replacement ion (NaCl) are 6 M, 60 mM and 80 mM, respectively. The cross-linker (BIS) concentrations are 5 mM (open circle), 100 mM (closed circle) and 200 mM (triangle). All isotherms were well fit by the Langmuir equation (Eq. (2)), as shown by the curves. Reproduced with permission from Wiley-VCH, Weinheim, Germany: Tanaka Memorial Symposium. Macromol. Symp. 2003 (Ref. [76]).

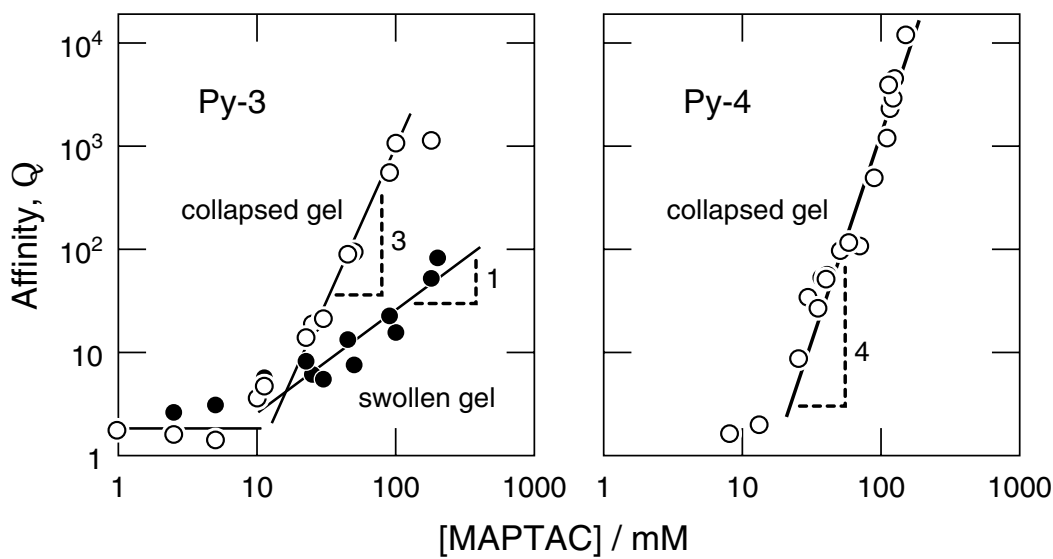

Figure 6: Adsorber concentration dependence of affinity for Py-3 and -4 adsorptions. Reproduced from Science 1999;286:1543–1545 (Ref. [21]).

## 3.2 Adsorber Concentration Dependence

$$Q \propto [\text{Ad}]^p \tag{26}$$

The variations of the affinity for Py-3 and Py-4 as a function of the MAPTAC concentration ([Ad]) are shown in Fig. 6. At higher MAPTAC concentrations in the collapsed state, both the log-log plots show a straight line, with slope three for Py-3 and with slope four for Py-4. In this experiment, an amount (100 mM) of salt was added to the system to sweep out the Donnan potential. The observed adsorption values suggest that the Donnan potential is not significant. Adsorption due to the Donnan potential was estimated to be at least three orders of magnitude smaller than the observed adsorption values. Furthermore, the non-specific adsorption of the Donnan potential should show a much weaker dependence on the MAPTAC concentration than the observed power laws of 3 or 4. Therefore, it is concluded that the power-law relationships observed here are due to three and four point adsorption, respectively. Adsorption sites are formed when three (or four) equivalent adsorbing molecules (MAPTA$^+$) gather to capture one Py-3 (or Py-4) molecule. The obtained power laws are explained well by the Tanaka equation discussed in the theoretical section.

At MAPTAC concentrations below 10 mM, the gel major component NIPA contributes more to the adsorption of pyranine (due to a hydrophobic interaction) than do the MAPTA$^+$ groups. The adsorption becomes independent of MAPTAC concentration and, consequently, the power-law exponent becomes zero.

In the swollen state, the log-log slope becomes one, indicating that MAPTA$^+$ adsorbs the target molecule with a single contact. Single point adsorption is favored because the MAPTA$^+$ monomers are well separated from one another. The slope return to 3 or 4 upon shrinking, indicating recovery of the multipoint binding sites.

These results suggest the following image. In the gel's collapsed state, the adsorber monomer units can be clustered to form sites in which a target molecule can bind with several adsorber monomer units simultaneously. If the gel is made to swell, the adsorber groups separate, and it becomes entropically unfavorable for the multipoint adsorption complex to assemble. In other words, the gels have a reversible adsorption ability which is controlled by the volume phase transition. In this sense, the binding sites in the gel behave like the active sites on a protein, as they can catch and release target molecules.

## 3.3 Salt Concentration Dependence

$$Q \propto [\text{Re}]^{-p} \tag{27}$$

In this section, the effect of external salt concentration ([Re]) on the binding of target molecules to the gel is presented. The gel system is similar to that used in the studies on adsorber dependence. In the adopted system, concentrations of the adsorber (MAPTAC) and the cross-linker (BIS) were fixed at 40 mM and 10 mM, respectively. The target molecules are adsorbed into the gel through an electrostatic interaction in which each target forms a complex with several adsorber monomers. Since the attraction is electrostatic, ions coexistent in the solution may affect the target molecule adsorption. Ions with the same charge sign as the target molecules should compete with the target molecules for binding with the adsorber monomers. Because these salt ions can replace target molecules in such binding, we refer to them as replacement molecules.

Figure 7 shows the affinity $Q$ of the collapsed gel for the various target molecules as a function of NaCl concentration. The affinities correlated very strongly with the NaCl concentration. For instance, in the case of Py-3, when the NaCl concentration was increased by a factor of three, the affinity decreased by more than two orders of magnitude. More importantly, the plots clearly show the linear relationships between ln([NaCl]) and ln($Q$) for the different target molecules. In the cases Py-1, Py-2, and Py-3, the log-log plot obviously has a slope of $-p_{\max}$, where $p_{\max}$ is the number of charged groups on the target molecule Py-$p_{\max}$. Py-4 follows similar behavior, though there was a discrepancy below ~40 mM of salt concentration. In this low salt regime, there was a negative deviation from the expected slope of $p_{\max} = -4$. The results suggest that the power relationships followed the proposed model in the whole range of salt concentrations observed for Py-1 to -3 and in the salt concentration higher than ~40 mM for Py-4. The exception of the lower salt concentration for Py-4 shows a limitation on the application of the model to the Py-4 adsorption system, which will be commented upon below.

In the previous section it was described how the Py-3 and Py-4 target molecules are captured in the collapsed gel with a number of contact points equal to their number of charges. The power law observed for the replacement molecule dependence is also determined by the charge number. The exponent, however, is negative, which suggests the following interpretation. The chloride ions of NaCl act as replacement ions for the charged groups on a target molecule. The target molecules have different numbers of contact points: one point for Py-1, two for Py-2, three for Py-3, and four for Py-4, respectively. The replacement ions substitute for the target molecules in binding to the adsorber, preventing adsorption of some of the targets. From an entropic point of view, the probability of inhibition should be proportional to the concentration of replacement ions raised to the power of the number of contact points. At the same time, from an energetic point of view, the binding energy of some $p$ salt ions should be proportional to $p$ as long as these ions are being adsorbed independently from each other. In other words, the power law, which has the form of a mass action law, indicates that adsorption centers (MAPTA$^+$ monomers) in the gel, even those involved in multipoint adsorption of a target molecule, are largely independent from each other.

It was also considered whether it is solely the ratio of adsorber to replacement concentration that determines the affinity, or whether the absolute value of the salt concentration plays some role [76]. For example, it is conceivable that there is some sort of coulombic screening that does not follow the mass action law. Figure 8 shows the variations of the affinity for Py-3 and Py-4 as a function of the ratio of the adsorber concentration [Ad] to the external salt concentration [Re]. The two variables, [Ad] and [Re], were tested in separate experiments. In one, the replacement ion concentration [Re] = [NaCl] = 80 mM was fixed while [Ad] was varied. In the other, [Ad] = [MAPTAC] = 40 mM was fixed while [Re] was varied. In both cases, [BIS] was 10 mM. The affinity curves from the two experiments are practically coincident. A slight deviation was observed for Py-4 at ratios larger than ~1, which corresponds to the low salt region of discrepancy observed in Fig. 7(d). This discrepancy is a limitation of the proposed model which may arise from the increasing effect of the Donnan potential at lower salt concentrations. Overall, the experimental evidence suggests that the absolute value of the salt concentration is not important to the affinity as long as the salt concentration is roughly equal to or larger than the adsorber concentration. In other words, the affinity is universally determined by the ratio of adsorber to replacement concentration, provided this ratio is smaller than 1.

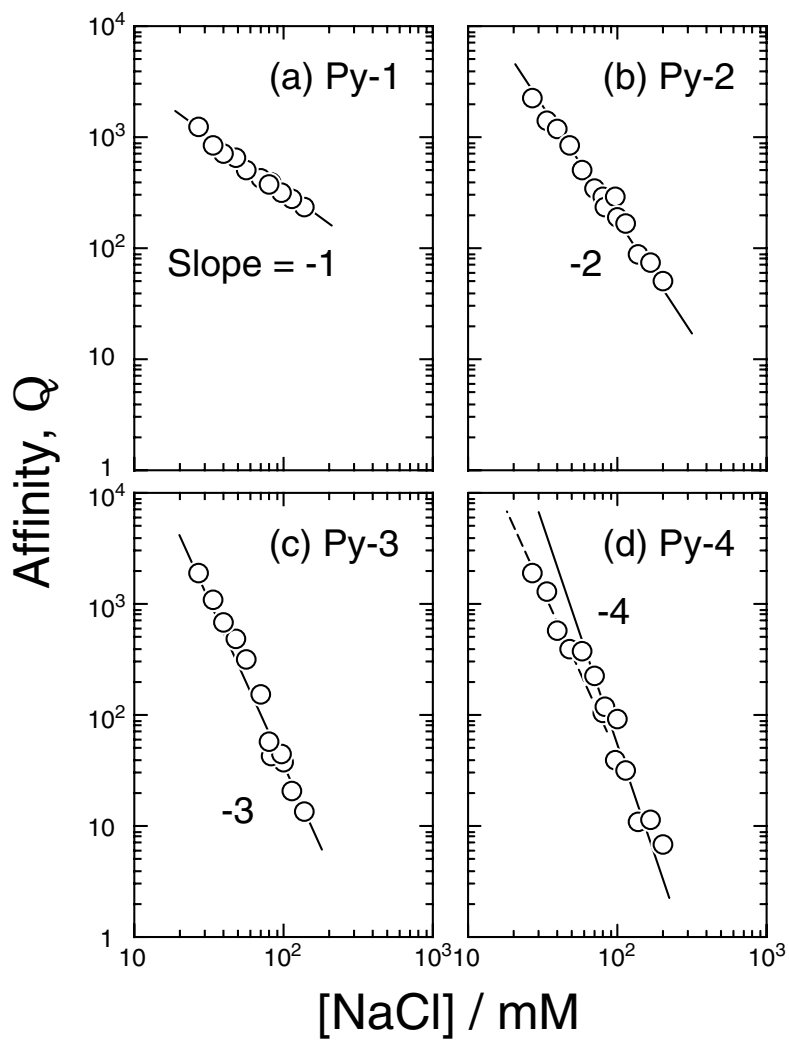

Figure 7: Replacement ion concentration dependence of affinity for Py-1 to -4 adsorptions. The solid line displays a slope with the value shown in each plot. The broken line in the plot (d) shows a slope of −3. A distinct tendency was observed in this plot, where the data slightly deviate from the trend line with a slope of −4 in the region of [Re] < [Ad]. See the text for a detail. Reproduced from J. Chem. Phys. 2001;115:1596–1600 (Ref. [75]).

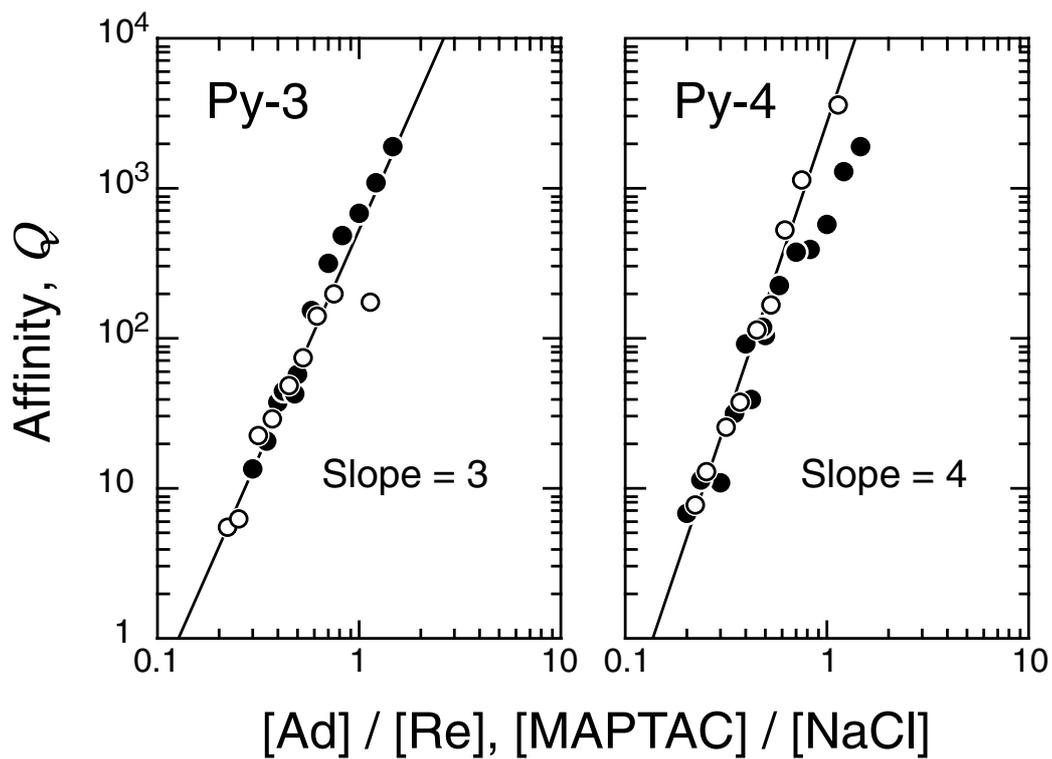

Figure 8: Dependence of the affinity on the ratio of the adsorber to the replacement ion for Py-3 and -4 adsorptions. The open and closed circles indicate experiments in which the adsorber or replacement ions were varied, respectively. Reproduced with permission from Wiley-VCH, Weinheim, Germany: Tanaka Memorial Symposium. Macromol. Symp. 2003 (Ref. [76]).

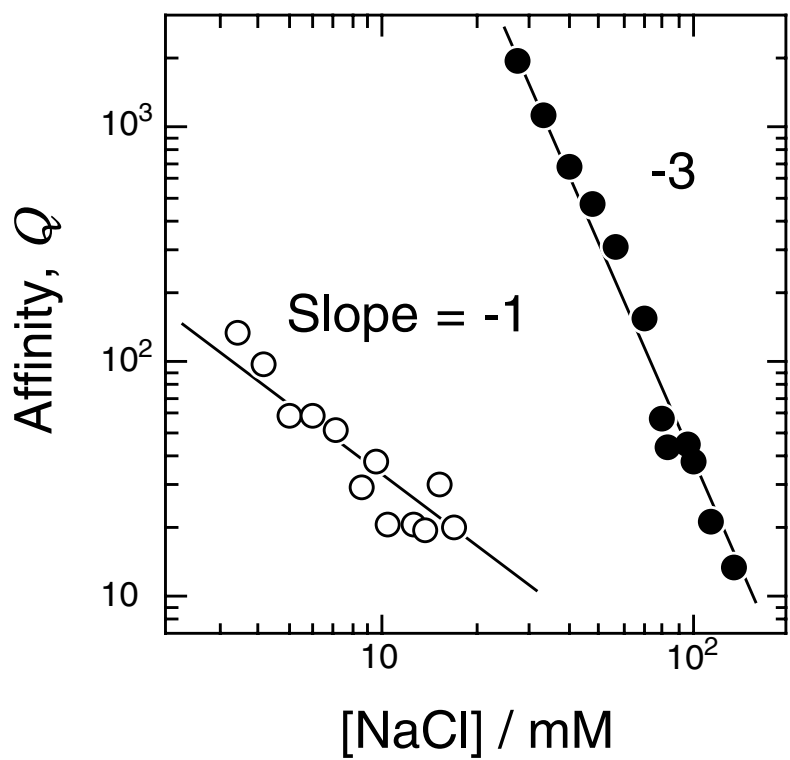

Figure 9: Log-log plot of the affinity *versus* the NaCl concentration for Py-3. The open circles denote the data for the swollen state at 25 °C, while the closed circles show the collapsed state at 60 °C. The NaCl concentrations were chosen such that the ratio of salt to adsorber covers the same ranges in both the swollen and collapsed states of the gel. Reproduced from J. Chem. Phys. 2001;115:1596–1600 (Ref. [75]).

Figure 9 is a comparison of the affinities of the swollen and collapsed gels for Py-3, as a function of NaCl concentration. Note that the salt concentration examined for the swollen gel was chosen so that the ratio [MAP-TAC]/[NaCl] covers the same range as for the collapsed gel experiment. The affinity decreased dramatically as the gel went from the collapsed state to the swollen state, even though the salt concentration was significantly lower in the swollen state. This affinity change across the phase transition was tested and found to be reversible. In addition, in the swollen state, the affinity decreased as the NaCl concentration was raised to 16.9 mM. The affinity could not be measured precisely at higher salt concentrations because adsorption of the target molecules became too weak.

Let us now compare the power-law relationship for the swollen gel and the collapsed gel. In the previous section, it was reported that the log-log plot between adsorber concentration and affinity gives a slope of $\sim 1$ for the swollen gel — independent of the charge number of the target — and a slope of $p$ for the collapsed gel [21]. This was interpreted to mean that the target is adsorbed with $p_{max}$ contact-points in the collapsed gel and 1 contact-point in the swollen gel. The data shown in Fig. 9 supports that interpretation.

We expected that the dependence of the affinity on the concentration of replacement ions should be controlled by the number of effective contact points. In the swollen gel, the power relationship between affinity and salt concentration was close to $-1$, although the data were somewhat scattered. In the collapsed gel, the power-law exponent was close to $-3$ for Py-3. This confirms that adsorption is single handed in the swollen state and multi-handed in the collapsed state.

The adsorption process is expected to be affected by the polymeric nature of the gel. Over short distances, the polymer chains are unconstrained. If the adsorbing monomers are close together (as in the collapsed state of the gel), the adsorbing monomers are free to gather into groups, which would allow multipoint adsorption. However, the adsorbers should not be able to diffuse beyond a certain length scale determined by the fraction of cross-linking monomers. In the gel's swollen state, the adsorber concentration is low, and the adsorbers may therefore be isolated from each other. Only single-handed adsorption may be possible in the swollen state of the gel. This can be inferred because of (1) the difference in the affinity power law between the swollen and collapsed states and (2) the much lower affinity in the swollen state, at all values of [MAPTAC]/[NaCl]. Another important aspect of the polymeric nature, the dependence of adsorption on the cross-linker concentration in the collapsed gel, will be described in the next section.

## 3.4 Cross-linker Concentration Dependence

$$Q \propto \exp\left(-(p-1)c\frac{[Xl]}{[Ad]^{2/3}}\right) \tag{28}$$

Previous papers have referred to the target molecules adsorbed through a multi-valent contact as 'gluons' [74] because, from a physical perspective, their essential property is to 'glue' together the adsorbing monomers on different polymer chains. The multipoint adsorption should lead to an entropy loss in the polymer chains bound by these gluons. The effect of this entropy loss, which is a function of the concentrations of the cross-linker and the adsorber, is to reduce the affinity of the gel for the gluon. For instance, several experimental studies have shown that the affinity of a heteropolymer gel involved in multiple contact adsorption decreases with its cross-linking degree. Hsein and Rorrer [90] showed an exponential decrease in calcium adsorption by chitosan as the extent of cross-linking increases. Eichenbaum et al. [91] found that, for alkali earth metal binding in methacrylic acid–co–acrylic acid micro gels, the cross-links prevent the carboxylic groups from achieving the same proximity as in a linear polymer, which affects the binding properties of the metals. In the case of a thermo-sensitive gel, the influence of the degree of cross-linking has also been shown [77, 78].

The volume phase transition of a NIPA gel induced by a stimulus is responsible for separating the adsorber monomers (methacrylic acid or MAPTAC groups) in the swollen state, which decreases their probability of coming close to each other to adsorb a multi-charged target. Consequently, in the swollen state, the affinity for the target increases slightly with the degree of cross-linking since the degree of swelling of the gel is reduced. In contrast, in the collapsed state an exponential decrease of the affinity with the cross-link density has been reported [77, 78]. This unfavorability for the affinity has been understood to be a 'frustration' to the mobility of the adsorbing sites [57, 77]. The frustration can also be viewed in terms of a 'flexibility' of the polymer chains. Flexibility of the polymer chains is critical for allowing the adsorber groups to come into proximity for the multipoint adsorption of a target molecule. A better understanding of this polymeric nature may allow us to move to the next stage, in which we control the frustration. In this section, substantial evidence for the frustration of target molecule adsorption is presented, and the validity of the naive model described above is discussed.

Figure 10 shows the dependence of the affinity $Q$ on the degree of cross-linking for several target molecules that can establish different numbers of contact points ($p$). For each target molecule, the affinities exponentially decrease as the cross-linker, BIS ([XI]), is increased. This effect is especially significant for the cases of contact numbers above two. As can be seen, the slope for the target with $p = 1$ is almost negligible, while for increasing $p$ the slope becomes larger. The initial slopes were well fit by an exponential function. This exponential nature can be understood by the discussion described above. Adsorption of the gluons by multiple contacts is affected by the required entropy loss of the polymer chains.

Let us consider the situation more quantitatively, from a microscopic point of view. Figure 11 shows a plot between the number of contact points and the exponential decay rate defined as

$$\text{slope} = -\frac{\partial (\ln Q)}{\partial \left(\frac{[BIS]}{[MAPTAC]^{2/3}}\right)}, \tag{29}$$

which is obtained from a fit of each plot in Fig. 10 to Eq. (8). The plot gives a linear relationship in which the slope is 0.32. The slope associates with the parameter $c$ in Eq. 1, though there is a correction factor because here the concentrations are reported in moles per liter, rather than particles per liter. This parameter $c$ can be used to calculate the persistence length $b = ma$ for the polymer chains, making use of the formula $c = \frac{2m}{[NIPA]b^2}$ from Eq. (7). As discussed in the theory section, $c$ may not provide the exact value of $b$, but should give a correct order of magnitude estimate. $b \sim 2.9$ nm is obtained with $c \sim 0.32$. The persistent length had been predicted to be around 2 nm, e.g. $\sim 10$ monomers ($m \sim 10$) of 2 Å length ($a = 2 \times 10^{-10}$ m) [92–94]. Thus, the obtained result is somewhat reasonable. In addition to this, analyzing the data for the adsorption of calcium ions in systems of NIPA/methacrylic acid gels [78] and NIPA/2–acrylamido–2–methylpropanesulfonic acid gels [80], it was found that the parameter $c$ had the values of 0.20 and 0.23, respectively. We have now verified that the theory predicts and explains well the exponential decay with concentration of the cross-linker. The cross-links and polymer connections create frustrations so that the adsorber groups (MAPTA$^+$) cannot lower the energy of the polymer by forming pairs, triplets, or groups of $p$ members to capture target molecules.

To test the effect of the adsorber concentration on the frustration (i.e. the factor $\frac{[XI]}{[Ad]^{2/3}}$) we examined the dependence of the affinity on the degree of cross-linking at various adsorber concentrations ([MAPTAC] = 30–60 mM) and a NaCl concentration fixed to 30 mM, for a target molecule (Py-4) that can establish four contact points ($p = 4$) [76]. The gels were prepared with 6 M NIPA. As was mentioned in the previous section, increasing the number of adsorber groups increases $Q$, since the saturation capacity of the gel becomes larger. According to Eq. 8, the slope of the cross-linker-affinity semi-log plot should be affected by the adsorber concentration. The results are shown in Fig. 12. We obtained four nearly parallel curves showing the exponential decrease of $Q$ with the cross-linking density for gels prepared at four different MAPTAC concentrations.

Figure 13 shows a log-log plot of the exponential decay rate for the data in Fig. 12 versus the MAPTAC concentration. In this case the exponential decay rate is redefined as $\frac{\partial (\ln Q)}{\partial ([BIS])}$. According to Eq. (8), the exponential decay rate should be proportional to a value $-3c[MAPTAC]^{-2/3}$ ($p = 4$). In Fig. 13 $\frac{\partial (\ln Q)}{\partial ([BIS])}$ decreases with $\ln([MAPTAC])$. A systematic decrease above $[MAPTAC] \geq 40$ mM and the predicted slope of $-2/3$, were observed, though the uncertainties are relatively large. The obtained trend in the plot does not necessarily conflict with the proposed model. The negative deviation at [MAPTAC] = 30 mM may be caused by a non-specific interaction with the NIPA main-chain, which would be more relevant at smaller values of $Q$.

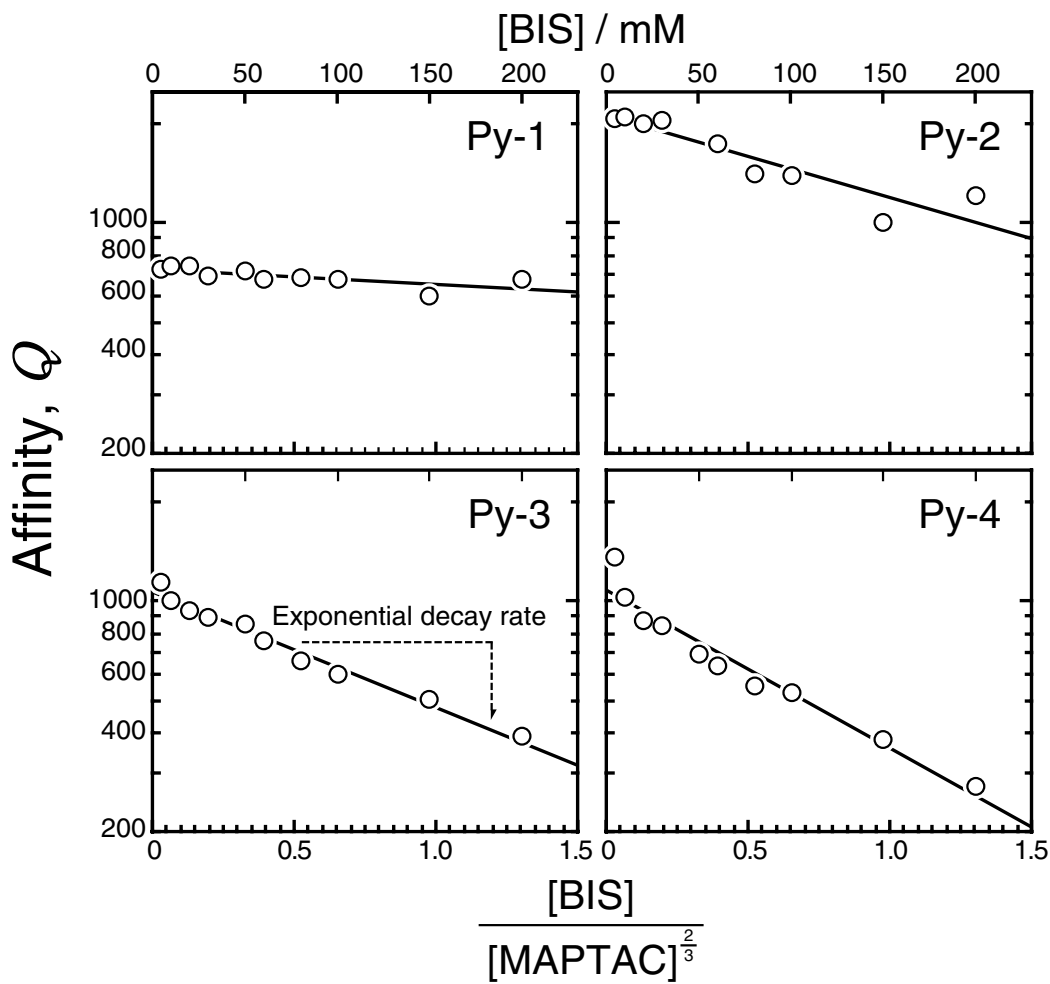

Figure 10: Series of semi-log plots for the affinity vs the BIS concentration for a collapsed gel at 60 °C. The concentrations of MAPTAC and NaCl are 60 mM and 80 mM, respectively. The lower horizontal label shows the ratio of BIS concentration to that of MAPTAC to the two-thirds power, while the upper label shows the corresponding BIS concentration. Reproduced with permission from Wiley-VCH, Weinheim, Germany: Tanaka Memorial Symposium. Macromol. Symp. 2003 (Ref. [76]).

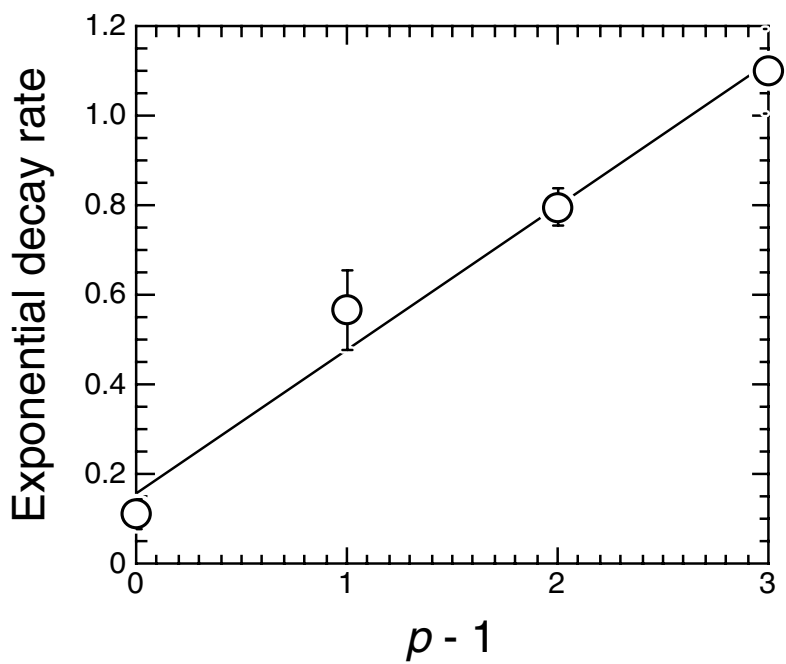

Figure 11: Plot of observed exponential decay rate vs $p - 1$, where $p$ is the number of contact points on the target molecule. The values of the decay rates and their uncertainness were obtained through a curve fitting using Eq. (8) on the data of Fig. 10. The plot shows a good linearity, with a slope of 0.32. The data were reproduced with permission from Wiley-VCH, Weinheim, Germany: Tanaka Memorial Symposium. Macromol. Symp. 2003 (Ref. [76]).

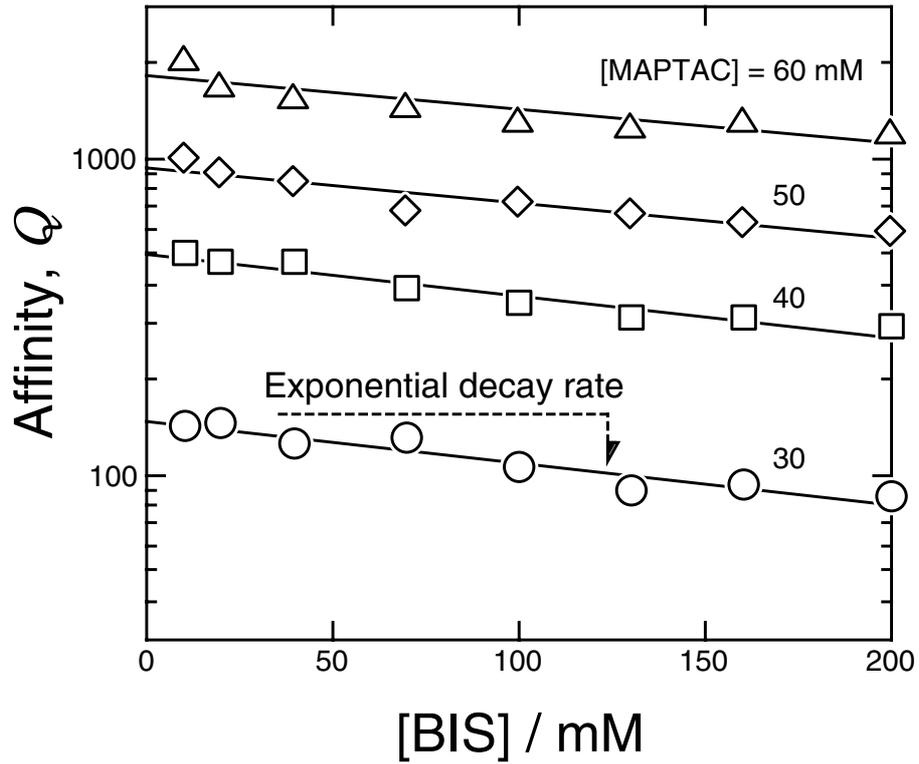

Figure 12: Dependence of the affinity on the degree of cross-linking at various adsorber concentrations, at a NaCl concentration of 30 mM for Py-4 adsorption by collapsed gels at 60 °C. The affinity decreases exponentially with cross-linker concentration, in accordance with Eq. (1). Reproduced with permission from Wiley-VCH, Weinheim, Germany: Tanaka Memorial Symposium. Macromol. Symp. 2003 (Ref. [76]).

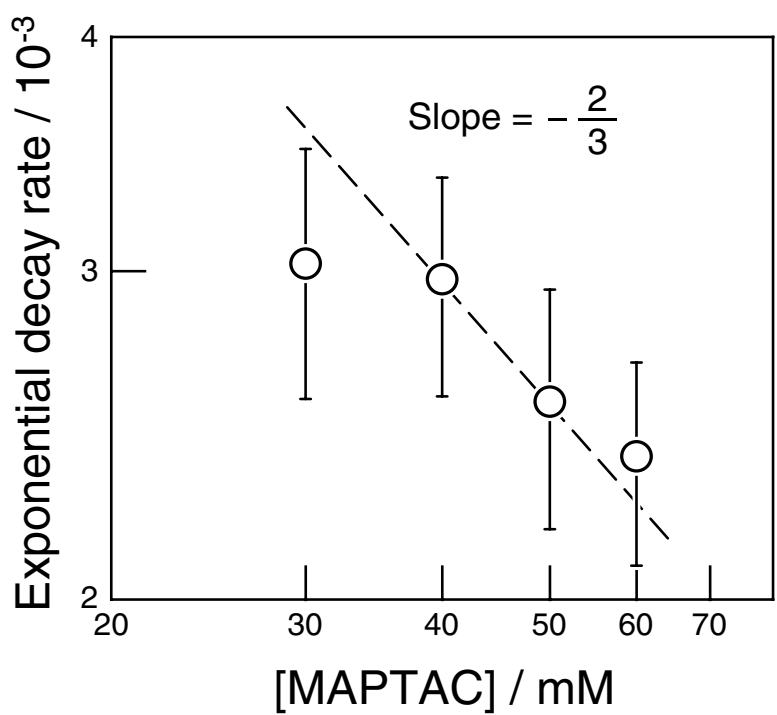

Figure 13: Double log plot of exponential decay rate versus the adsorber concentration. The uncertainness was obtained from a curve fitting using Eq. (8) on the data of Fig. 12. The data were reproduced with permission from Wiley-VCH, Weinheim, Germany: Tanaka Memorial Symposium. Macromol. Symp. 2003 (Ref. [76]).

# 4   Applications to Imprinted Gel Systems

A polymer gel can perform some of the fundamental functions of proteins [45,74]. Like proteins, a heteropolymer gel can exist in four thermodynamic phases: (1) Swollen and fluctuating, (2) Shrunken and fluctuating, (3) Shrunken and frozen in a degenerate conformation, and (4) Shrunken and frozen in the global minimum energy conformation. The order parameter that describes the phase transition between the first and the second phases is the polymer density, or equivalently, the swelling ratio of the gel. The third and fourth phases are distinguished by another order parameter: the overlap between the frozen conformation and the minimum energy conformation. In the third phase, the frozen conformation is random. The third phase is experimentally observed in gels through a multiple phase transition in a polyampholyte gel [15]. In the fourth phase, the frozen conformation is equivalent to that of the global energy minimum. Proteins in this fourth phase take on a specific conformation, which may be capable of performing catalysis, molecular recognition, or many other activities. Tanaka and colleagues strove to recreate such a fourth phase in gels by designing a low energy conformation, and then testing whether the gel could be made to reversibly collapse into this 'memorized' conformation.

According to developments in the statistical mechanics of polymers, the memory of conformation by flexible polymer chains has several requisites [59]. First, the polymer must be a heteropolymer, i.e. there should be more than one monomer species, so that some conformations are energetically more favorable than others. Second, there must be frustrations which hinder a typical polymer sequence from being able to freeze to its lowest energy conformation (as considered in absence of the frustration). Such frustrations may be due to the interplay of chain connectivity and excluded volume or may be created by cross-links. For example, a cross-linked polymer chain will not freeze into the same conformation as the uncross-linked polymer chain, at least for most polymer sequences. Such frustrations are possible when the polymer is in its condensed state. Third, the sequence of monomers must be selected as to minimize these frustrations, i.e. a particular polymer sequence should be designed such that the frustrating constraints do not hinder the polymer from reaching its lowest energy conformation. These three conditions allow the polymer to have a global free energy minimum at one designed conformation.

The experimental gel systems described in previous sections satisfy the first two conditions, and can be engineered to satisfy the third. The adsorber monomers and main component monomers provide heterogeneity in interaction energies, since adsorber interactions are favorably mediated by 'gluon' target molecules [74]. Frustrations to the achievement of the global energy minimum exist due to the cross-links in the gel, as well as chain connectivity. To meet the third condition, that is, the minimization of the frustration, the 'imprinting technique' [63–67] was adopted. In this section, we discuss the early experimental successes [77–84] of the imprinting method and the application of the Tanaka equation to imprinted gels.

Two approaches were investigated to achieve elements of conformational memory in gels: (1) cross-linking of a polymer dispersion in the presence of the target molecule (*two-step* imprinting or *post-cross-linking*), or (2) cross-linking a monomer solution while the monomers polymerize (*one-step* imprinting). The first approach was studied in heteropolymer gels consisting of NIPA as the major monomer component sensitive to stimuli and charged monomers as the component able to capture target molecules [77]. The polymer networks were randomly copolymerized with a small quantity of permanent cross-links and thiol groups (–SH). The gels were then further cross-linked by connecting thiol group pairs into disulfide bonds (–S–S–). Because this second step takes place after the initial polymerization of the gel, we refer to it as *post-cross-linking*. These post-cross-links were still in very low concentrations in the range 0.1–3 mol%, under which condition the polymers could still freely swell and shrink to undergo the volume phase transition [77]. Gels that were post-cross-linked while all charged monomers were in complex formation with the target molecules (post-imprinting) showed higher affinity for the target than the gels that were randomly post-cross-linked. The post-imprinted gels showed that assemblies of the charged monomers were memorized after swelling and re-collapsing (Fig. 14) and that the frustrations were partially minimized.

However, this *post-cross-linking* approach has a fundamental drawback. Before the *post-cross-linking*, the sequence of the components has already been determined and randomly quenched. The minimization of the frustration is allowed only in the freedom of finding best partners among –SH groups. For this reason, the imprinting of conformational memory by *post-cross-linking* can give only a partial success. Ideally, the entire sequence of all monomers must be chosen so that the system will be in its global energy minimum. The complete minimization of the frustration can be achieved by polymerizing monomers while they self-organize in space at a low-energy spatial arrangement [64, 74, 95]. Based on these ideas, the second approach was pursued, described as follows.

The first successes of the complete minimization of the frustration in gels by *one-step* imprinting were reported by Alvarez-Lorenzo et al. [78, 79]. In this approach, to control the sequence formation, the monomers are allowed to equilibrate in the presence of a target molecule. These monomers are then polymerized with a cross-linker, producing a plastic. It is hoped that upon removal of the template species, cavities are formed in the polymer matrix and that these holes have memorized the spatial features and bonding preferences of the template. The choice of functional monomers together with the spatial arrangement of functional groups are two of the main factors responsible for specificity and reversibility of molecular recognition.

A control group of random gels was prepared by random polymerization of NIPA and small amounts of methacrylic acid and BIS in dioxane at $60\,^\circ$C. Imprinted gels were prepared by incorporating lead-methacrylate where lead ions were used as divalent ion templates for creating two-point adsorption sites. The imprinted gels were then swollen, and the lead ions were washed out, freeing the methacrylic groups from each other.

The measurements of the affinity suggested that multipoint adsorption occurs for both gels in the collapsed state, but that in the imprinted gel, the multipoint adsorption is due to memorized binding sites [78]. For both gels, there was a strong affinity for the divalent ions in the collapsed state. When the gels swell, the adsorption sites are disrupted, leading to lower affinity and single-handed adsorption, as shown in Fig. 15. After recollapse, two-point adsorption is recovered in the random gels with a power law of $Q \sim [Ad]^2$ (Fig. 16). For imprinted gels, however, the affinity is stronger and the power law is $Q \sim [Ad]^1$. Meanwhile, the saturation value $S$ obeys $S \sim [Ad]/2$ for both gels, from which it is concluded that multipoint adsorption occurs for both gels in the collapsed state. The difference in the power dependence of $Q$ is due to different [Ad] dependence for $K$, the affinity per adsorption site. $K$ is actually independent of [Ad] for imprinted gels. A qualitative explanation of these results is as follows. Since an increase in [Ad] does not improve the affinity at the adsorption sites in an imprinted gel, we surmise that the adsorbers are ordered such that they already form sites with their unique partners. Hence, the adsorbers are ordered to give the highest possible $K$, at all values of [Ad].

The affinity of the imprinted gels did not change with an increase of the cross-linking density, while the affinity decreased for random gels (Fig. 17). As seen in the Tanaka equation, [Xl] is part of the entropic term due to polymeric constraints. The complete disappearance of [Xl] dependence implies the minimization of the frustration due to the cross-links and chain connectivity. Thus requirements one, two, and three necessary for conformational memory have been achieved. It is concluded that that the sequence of monomers can be selected

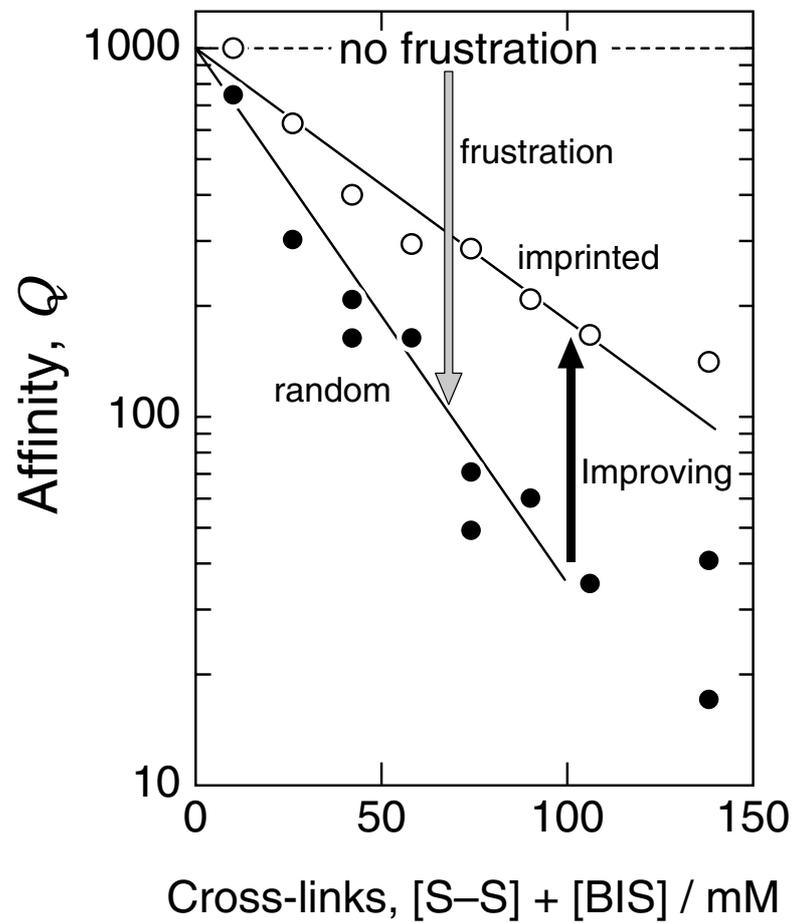

Figure 14: Imprinting effect using post-cross-linking. The decay in the affinity due to frustration is observed for the random gel (closed circle). The decay rate of the imprinted gel is substantially improved by the post-imprinting (open circle). Reproduced from Phys. Rev. Lett. 2000;85:5000–5003 (Ref. [77]). Copyright(2000) by the American Physical Society.

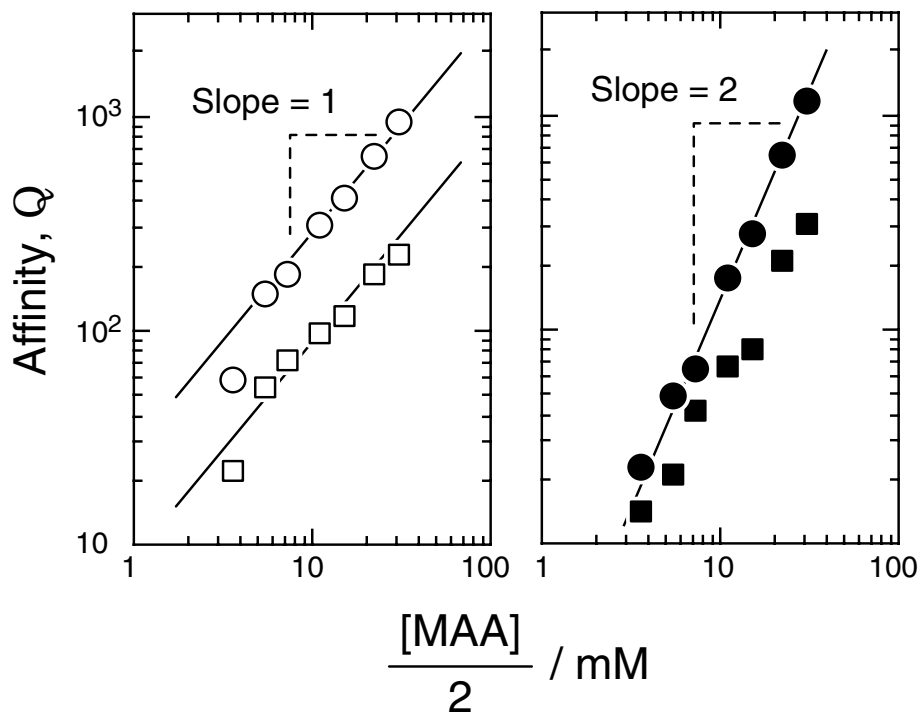

Figure 15: Imprinting effect using *one-step* process. The left and right hand plots display the affinities for the imprinted and random gels, respectively. The circle and square represent for the collapsed and swollen states. Reproduced with permission from Macromolecules 2000;33:8693–8697 (Ref. [78]). Copyright(2000) the American Chemical Society.

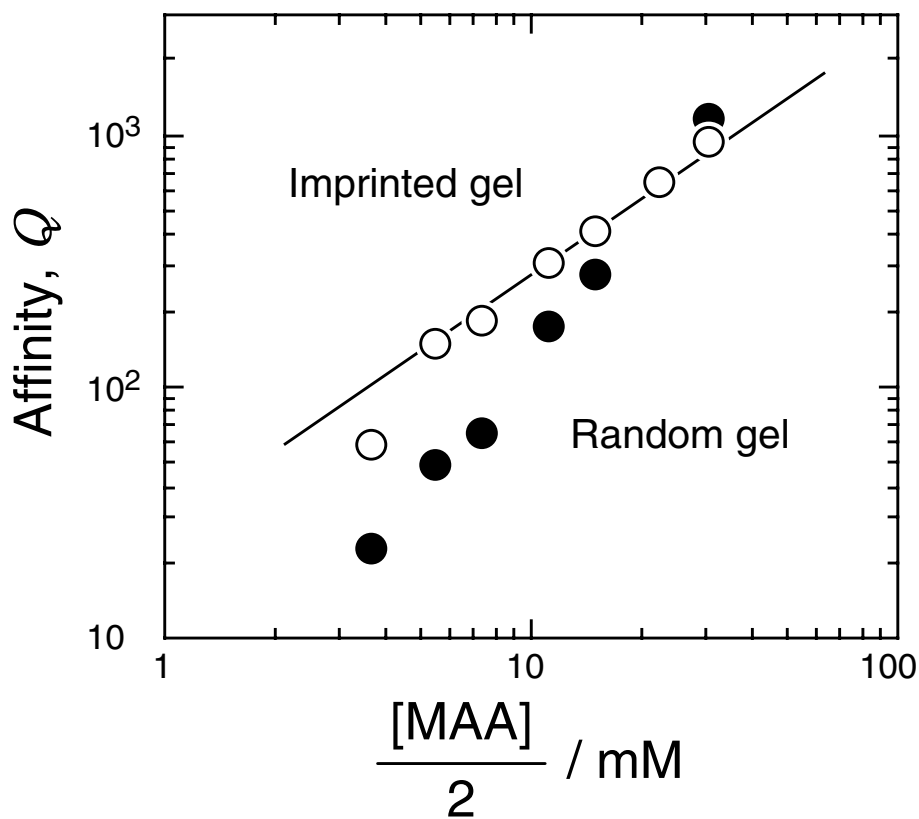

Figure 16: Imprinting effect using *one-step* process. Comparison between the affinities of the imprinted and random gels. The significant increase of the affinity is observed for the imprinted gel, and the cross point of the respective trend lines falls on the corresponding concentration of the cross-linker (40 mM/2), suggesting the improvement by the imprinting is effective with the frustration. Reproduced with permission from Macro-molecules 2000;33:8693–8697 (Ref. [78]). Copyright(2000) the American Chemical Society.

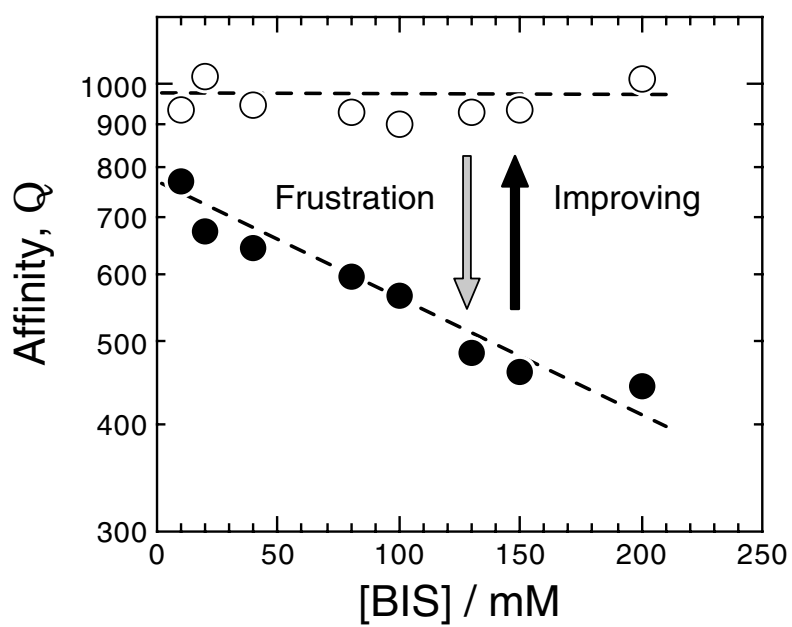

Figure 17: Imprinting effect using *one-step* process. No frustration is observed in the cross-linker concentration dependence of the affinity for the imprinted gel. Reproduced with permission from Macromolecules 2000;33:8693–8697 (Ref. [78]). Copyright(2000) the American Chemical Society.

to create molecular-scale conformational memory and minimize the frustration in heteropolymer gels.

Having demonstrated the possibility of conformational memory in imprinted gels, some characterizing tests were performed. Ando recently examined how salt concentration affects the affinity of the imprinted gel [82]. The effects of [Re] on the adsorption of calcium ions by the same lead-methacrylate imprinted gel system were studied by varying the concentration of NaCl in the target solutions [82]. Gels were prepared with 6 M NIPA, 100 mM BIS, and 50 mM methacrylic acid for a random gel and 25 mM lead-methacrylate for imprinted. According to the Tanaka equation, $Q$ should be identical for both random and imprinted gels, since the replacement molecule dependence is due to interactions of the salt with the target molecule. The preliminary results showed that the dependencies of the affinity $Q$ on the replacement ion concentration are similar for random and imprinted gels. These results roughly agree with the power laws of $Q \sim [\text{Re}]^{-1}$ for the swollen state and $Q \sim [\text{Re}]^{-2}$ for the collapsed state observed by Watanabe et al. [75] in random gels, though there were uncertainties due to possible incomplete incorporation of lead-methacrylate or incomplete removal of lead in the imprinted gels. These studies further suggest that $S$ may be independent of [Re] for imprinted gels, though more experiments are necessary for verification.

The success of imprinting depends strongly on the stability and solubility of the complexes template/functional monomers formed before polymerization; therefore the choice of solvent used in the polymerization process is critical. The positive results of Alvarez-Lorenzo et al. can be attributed to the fact that lead and two methacrylate molecules remain bound during polymerization. If the molar ratio in the complex is not appropriate or if the complex dissociates to some extent during polymerization, the functional monomers will be far apart from both the template and each other, and imprinting will be thwarted. The effects of different solvents have been observed in imprinted gels prepared with divalent salts of methacrylic monomers [79].

Another problem is incomplete removal of the template after polymerization, which leads to an impure gel. The subsequent release of the template from the network can interfere with adsorption studies. Recent efforts showed the possibilities of using adsorbing monomers directly bonded to each other prior to polymerization, which avoids the use of the template polymerization technique [80, 81]. Each adsorbing monomer can be broken after polymerization to obtain pairs of ionic groups with the same charge. Since the members of each pair are close together, they can capture target molecules through multipoint ionic interactions. The adsorption process was found to be independent of the cross-linking density and the entropic frustrations were completely resolved.

## 5   Conclusion

Heteropolymer gels are an excellent system for the study of molecular adsorption of target molecules. In this review, we have formulated a theory for the adsorption of target molecules into a random heteropolymer gel and have verified the theory by experiments. The gel has been modeled as a set of adsorbing monomers connected by gaussian chains to fixed cross-linking points. The theory can be summarized by the Tanaka equation, which predicts power-law dependence of the gel affinity with regard to the adsorbing monomer concentration, salt concentration, and cross-linker concentration. The predicted affinity dependence on these quantities has been assessed through a large series of experiments which we have summarized. By adjusting the concentrations of these variables, one can control the affinity of a heteropolymer gel by several independent, but well-characterized means.

Furthermore, and more importantly, it has been established that for gels of the appropriate composition, adsorption occurs through multiple contacts in the collapsed state, but through single contacts in the swollen state. This multipoint adsorption makes the collapsed state affinity significantly larger than that in the swollen state. The abrupt change in affinity during the gel volume phase transition allows one to turn gel adsorption "on" or "off," by using any of the well-established means of forcing a phase transition.

These observations on random gels provide the basis for imprinted gels, which can be designed to have even higher affinity and specificity for target molecules. Like random gels, imprinted gels can be turned "on" or "off," but hold the further promise of specific molecular recognition, through the minimization of frustration in the polymerization process. We have provided early experimental results for successful imprinting in heteropolymer gels.

It is expected that the Tanaka equation, and its modifications by imprinting, will be applicable to any heteropolymer gel adsorbing a target molecule, and not be limited to the chemical components we have mentioned here. In this review we have summarized target molecule adsorption in several different gel systems with various components. By embedding other types of interactions into the gel, even higher selectivity should be possible, and for a variety of systems. The diversity of interactions offer the potential for cooperativity between repulsive and attractive effects, an essential property for the volume phase transition of a polymer gel.

This paper represents a summary of several years of research on imprinting and adsorption in which a number of promising discoveries have been made. Still, much work remains to be done for the long-term goal of achieving enzyme-like specificity, as was envisioned by Professor Tanaka and the many others in this field.

# References

[1] Dušek K., Patterson D. Transition in swollen polymer networks induced by intramolecular condensation. J. Polym. Sci: Part A2, 1968;6 1209–1216.

[2] Tanaka T. Collapse of gels and the critical endpoint. Phys. Rev. Lett. 1978;40 820–823.

[3] Tanaka T. Phase transitions in gels and a single polymer. Polymer 1979;20 1404–1412.

[4] Tanaka T., Swislow G., Ohmine I. Phase separation and gelation in gelatin gels. Phys. Rev. Lett. 1979;42 1556–1559.

[5] Hockberg A., Tanaka T., Nicoli D. Spinodal line and critical point of an acrylamide gel. Phys. Rev. Lett. 1979;43 217–219.

[6] Tanaka T., Fillmore D., Sun S-T., Nishio I., Swislow G., Shah A. Phase transition in ionic gels. Phys. Rev. Lett. 1980;45 1636–1639.

[7] Tanaka T., Nishio I., Sun S.-T., Ueno-Nishio S. Collapse of gels in an electric field. Science 1982;218 467–469.

[8] Hirokawa Y., Tanaka T., Katayama S. Effects of network structure on the phase transition of acrylamide-sodium acrylate copolymer gels. In: Marchall K. C. editor. Dahlem workshop reports life sciences research report 31 microbial adhesion and aggregation. Berlin: Springer-Verlag, 1984. p.177–188.

[9] Tanaka T., Sato E., Hirokawa Y., Hirotsu S., Peetermans J. Critical kinetics of volume phase transition of gels. Phys. Rev. Lett. 1985;55 2455–2458.

[10] Ilmain F., Tanaka T., Kokufuta E. Volume transition in a gel driven by hydrogen bonding. Nature 1991;349 400–401.

[11] Tanaka T., Sun S-T., Hirokawa Y., Katayama S., Kucera J., Hirose Y., Amiya T. Mechanical instability of gels at the phase transition. Nature 1987;325 796–798.

[12] Suzuki A., Tanaka T. Phase transition in polymer gels induced by visible light. Nature 1990;346 345–347.

[13] Tokita M., Tanaka T. Reversible decrease of gel-solvent friction. Science 1991;253 1121–1123.

[14] Kokufuta E., Zhang Y-Q., Tanaka T. Saccharide-sensitive phase transition of a lectin-loaded gel. Nature 1991;351 302–304.

[15] Annaka M., Tanaka T. Multiple phases of polymer gels. Nature 1992;355 430–432.

[16] Sato-Matsuo E., Tanaka T. Patterns in shrinking gels. Nature 1992;358 482–485.

[17] Zhang Y-Q., Tanaka T., Shibayama M. Super-absorbency and phase transition of gels in physiological salt solutions. Nature 1992;360 142–144.

[18] Li Y., Tanaka T. Phase transition of gels. Annu. Rev. Mater. Sci. 1992;22 243–277.

[19] Tanaka T., Annaka M., Ilmain F., Ishii K., Kokufuta E., Suzuki A., Tokita M. Phase transitions of gels. In: Karalis T. K. editor. NATO ASI series, H64, mechanics of swelling. Berlin: Springer-Verlag, 1992. p.683–703.

[20] Shibayama M., Tanaka T. Volume phase transition and related phenomena of polymer gels. In: Dušek K. editor. Advances in polymer science, responsive gels: Volume transitions I;109. Belrin: Springer-Verlag, 1993. p.1–62.

[21] Oya T., Enoki T., Grosberg A. Yu., Masamune S., Sakiyama T., Takeoka Y., Tanaka K., Wang G., Yilmaz Y., Feld M. S., Dasari R., Tanaka T. Reversible molecular adsorption based on multiple-point interaction by shrinkable gels. Science 1999;286 1543–1545.

[22] Hirokawa Y., Tanaka T., Sato-Matsuo E. Volume phase transition in a nonionic gel. J. Chem. Phys. 1984;81 6379–6380; *ibid* 1992;96 8641.

[23] Hirotsu S., Hirokawa Y., Tanaka T. Volume-phase transitions of ionized N-isopropylacrylamide gels. J. Chem. Phys. 1987;87 1392–1395.

[24] Otake K., Inomata H., Konno M., Saito S. A new model for the thermally induced volume phase-transition of gels. Journal of Chemical Physics 1989; 91(2) 1345–1350.

[25] Bae Y. H., Okano T., Kim S. W. Temperature-dependence of swelling of cross-linked poly(N,N'-alkyl substituted acrylamides) in water. Journal of Polymer Science Part B-Polymer Physics 1990; 28(6) 923–936.

[26] Seida Y., Nakano Y., Ichida H. Surface-properties of temperature-sensitive N-isopropylacrylamide-copolymer gels. Kagaku Kogaku Ronbunshu 1992; 18(3) 346–352.

[27] Gehrke S. H. Synthesis, equilibrium swelling, kinetics, permeability and applications of environmentally responsive gels. In *Advances in Polymer Science, Responsive Gels: Volume Transitions II*, vol. 110, ed. K. Dušek, Springer-Verlag, Belrin, 1993, pp. 81–144.

[28] Ilavský M. Phase-transition in swollen gels. 2. Effect of charge concentration on the collapse and mechanical-behavior of polyacrylamide networks. Macromolecules 1982;15(3) 782–788.

[29] Katayama S., Ohata A. Phase transition of a cationic gel. Macromolecules 1985;18(12) 2781–2782.

[30] Myoga A., Katayama S. Volume phase transition of amphoteric gel. Polymer Prep. Japan 1987;36 2852–2853.

[31] Katayama S., Myoga A., Akahori Y. Swelling behavior of amphoteric gel and the volume phase transition. J. Chem. Phys. 1992;96 4698–4701.

[32] Amiya T., Tanaka T. Phase transitions in cross-linked gels of natural polymers. Macromolcules 1987;20 1162–1164.

[33] Siegel R. A., Firestone B. A. pH-dependent equilibrium swelling properties of hydrophobic poly-electrolyte copolymer gels. Macromolecules 1988;21 3254–3259.

[34] Siegel R. A., Faramalsian M., Firestone B. A., Moxley B. C. pH-controlled release from hydrophobic poly-electrolyte copolymer hydrogels. J. Controlled Release 1988;8 179–182.

[35] Gehrke S. H., Cussler E. L. Mass transfer in pH-sensitive hydrogels. Chem. Eng. Sci. 1989; 44 559–566.

[36] Katayama S., Myoga A., Akahori Y. Swelling behaviors of amphoteric gels and the dissociation mechanisms. Polym. Bull. 1992;28 227–233.

[37] Ohmine I. Tanaka T. Salt effects on the phase transitions of polymer gels. J. Chem. Phys. 1982;77 5725–5729.

[38] Rička J., Tanaka T. Swelling of ionic gels: quantitative performance of the donnan theory. Macromolecules 1984;17 2916–2921; Phase Transition in Ionic Gels Induced by Copper Complexation. Macromolecules 1985;18 83–85.

[39] Osada Y. Conversion of chemical into mechanical energy by synthetic-polymers (chemomechanical systems). Adv. Polym. Sci. 1987;82 1–46.

[40] Osada Y., Umezawa K., Yamauchi A. Oscillation of electrical-current in water-swollen poly-electrolyte gels. Macromol. Chem. Phys. 1988;189 597–605.

[41] Gianetti G., Hirose Y., Hirokawa Y., Tanaka T. Effects of D.C. electric fields on gels. In: Carter F. L., Siatkowski R. E., Wohltjen H. editors. Molecular Electronic Devices. London: Elsevier, 1988. p.369–380.

[42] Kurauchi T., Shiga T., Hirose Y., Okada A. Deformation behaviors of polymer gels in electric-field. In: DeRossi, D., Kajiwara K., Osada Y., Yamauchi A. editors. Polymer Gels: Fundamentals and biomedical applications. New York: Plenum Press, 1991. p.237–246.

[43] Mamada A., Tanaka T., Kungwatchakun D., Irie M. Photoinduced phase transition of gels. Macromolecules 1990;23 1517–1519.

[44] Kataoka K., Miyazaki H., Bunya M., Okano T., Sakurai Y. Totally synthetic polymer gels responding to external glucose concentration: their preparation and application to on-off regulation of insulin release. J. Am. Chem. Soc. 1998;120 12694–12695.

[45] Wang G., Kuroda K., Enoki T., Grosberg A. Yu., Masamune S., Oya T., Takeoka Y., Tanaka T. Gel catalysis that switch on and off. Proc. Natl. Acad. Sci. 2000;97 9861–9864.

[46] Osada Y., Okuzaki H., Hori H. A polymer gel with electrically driven motility. Nature 1992; 355(6357) 242–244.

[47] Peppas N. A., Langer R. New challenges in biomaterials. Science 1994;263(5154) 1715–1720.

[48] Hu, Z., Zhang, X., Li Y., Synthesis and application of modulated polymer gels. Science 1995;269 525–527.

[49] Yoshida R., Uchida K., Kaneko Y., Sakai K., Kikuchi A., Sakurai Y., Okano T. Comb-type grafted hydrogels with rapid de-swelling response to temperature-changes. Nature 1995;374(6519) 240–242.

[50] Yoshida R., Takahashi T., Yamaguchi T., Ichijo H. Self-oscillating gel. Journal of the American Chemical Society 1996;118(21) 5134–5135.

[51] Yoshida R., Takei K., Yamaguchi T. Self-beating motion of gels and modulation of oscillation rhythm synchronized with organic acid. Macromolecules 2003;36(6) 1759–1761.

[52] Holtz J. H., Asher S. A. Polymerized colloidal crystal hydrogel films as intelligent chemical sensing materials. Nature 1997;389(6653) 829–832.

[53] Calvert P. Electroactive polymer gels. In: Bar-Cohen Y. editor. Electroactive polymer (EAP) actuators as artificial muscles–reality, potential and challenges. Bellingham: SPIE Press, 2001. p.123–138.

[54] Takeoka Y., Watanabe M. Polymer gels that memorize structures of mesoscopically sized templates. Dynamic and optical nature of periodic ordered mesoporous chemical gels. Langmuir 2002;18(16) 5977–5980.

[55] Kikuchi A., Okano T. Intelligent thermoresponsive polymeric stationary phases for aqueous chromatography of biological compounds. Progress in Polymer Science 2002;27(6) 1165–1193.

[56] Alvarez-Lorenzo C., Hiratani H., Gómez-Amoza J.L., Martínez-Pacheco R., Souto C., Concheiro A. Soft contact lenses capable of sustained delivery of timolol. J. Pharm. Sci. 2002; 91(10) 2182–2192.

[57] Takeoka Y., Berker A. N., Du R., Enoki T., Grosberg A., Kardar M., Oya T., Tanaka K., Wang G., Yu X., Tanaka T. First order phase transition and evidence for frustration in polyampholytic gels. Phys. Rev. Lett. 1999;82 4863–4865.

[58] Pande V. S., Grosberg A. Yu., Tanaka T. Is heteropolymer freezing well described by the random energy model? Phys. Rev. Lett. 1996;76 3987–3990.

[59] Pande V. S., Grosberg A. Yu., Tanaka T. Heteropolymer freezing and design: towards physical models of protein folding. Rev. Mod. Phys. 2000;72 259–314.

[60] Shakhnovich E. I., Gutin A. M. A new approach to the design of stable proteins. Protein Engineering 1993;6 793–800.

[61] Frauenfelder H., Sligar S. G., Wolynes P. G. The energy landscapes and motions of proteins. Science 1991;254(5038) 1598–1603.

[62] Wolynes P. G., Onuchic J. N., Thirumalai D. Navigating the folding routes. Science 1995;267(5204) 1619–1620.

[63] Wulff G., Sarhar A., Zabrocki K. Enzyme-analogue built polymers and their use for the resolution of racemates Tetrahed. Lett. 1973;44 4329–4332.

[64] Wulff G. Molecular imprinting in cross-linked materials with the aid of molecular templates – a way towards artificial antibodies. Angew. Chem.-Int. Edit. 1995;34 1812–1832.

[65] Wulff G. Enzyme-like catalysis by molecularly imprinted polymers. Chemical Reviews 2002;102 1–28.

[66] Bartsch R. A., Maeda M. editors. Molecular and Ionic Recognition with Imprinted Polymers, ACS Symposium Series; 703. Washington DC.: American Chemical Society, 1998.

[67] Ramström O., Ansell R. J. Molecular imprinting technology: challenges and prospect for the future. Chirality 1998;10 195–209.

[68] Watanabe M., Akahoshi T., Tabata Y., Nakayama, D. Molecular specific swelling change of hydrogels in accordance with the concentration of guest molecules. J. Am. Chem. Soc. 1998;120(22) 5577–5578.

[69] Allender C. J., Keith B., Heard C. M. Molecularly-imprinted polymers. Preparation, biomedical applications and technical challenges. In: King F. D., Oxford A. W. editors. Progress in Medicinal Chemistry, Vol. 36. Amsterdam: Elsevier Science, 1999. p.235–291.

[70] Miyata T., Asami N., Uragami T. A reversibly antigen-responsive hydrogel. Nature 1999;399(6738) 766–769.

[71] Miyata T., Uragami T., Nakamae K. Biomolecule-sensitive hydrogels. Advanced Drug Delivery Reviews 2002;54(1) 79–98.

[72] Asanuma H., Hishiya T., Komiyama M. Tailor-made receptors by molecular imprinting. Adv. Mater. 2000;12(14) 1019–1030.

[73] Peppas N. A., Huang Y. Polymers and gels as molecular recognition agents. Pharmaceutical Research 2002;19(5) 578–587.

[74] Tanaka T., Enoki T., Grosberg A. Y., Masamune S., Oya T., Takeoka Y., Tanaka K., Wang C., Wang G. Reversible molecular adsorption as a tool to observe freezing and to perform design of heteropolymer gels. Ber. Bunsenges. Phys. Chem. 1998;102 1529–1533.

[75] Watanabe T., Ito K., Alvarez-Lorenzo C., Grosberg A. Yu., Tanaka T. Affinity control through salt concentration on multiple contact adsorption into heteropolymer gel. J. Chem. Phys. 2001;115 1596–1600.

[76] Ito K., Chuang J., Alvarez-Lorenzo C., Watanabe T., Ando N., Grosberg A. Yu. Multiple contact adsorption of target molecules. Macromol. Symp. 2003; in press.

[77] Enoki T., Tanaka K., Watanabe T., Oya T., Sakiyama T., Takeoka Y., Ito K., Wang G., Annaka M., Hara K., Du R., Chuang J., Wasserman K., Grosberg A. Y., Masamune S., Tanaka T. Frustrations in polymer conformation in gels and their minimization through molecular imprinting. Phys. Rev. Lett. 2000;85 5000–5003.

[78] Alvarez-Lorenzo C., Guney O., Oya T., Sakai Y., Kobayashi M., Enoki T., Takeoka Y., Ishibashi T., Kuroda K., Tanaka K., Wang G., Grosberg A. Y., Masamune S., Tanaka T. Polymer gels that memorize elements of molecular conformation. Macromolecules 2000;33 8693–8697.

[79] Alvarez-Lorenzo C., Guney O., Oya T., Sakai Y., Kobayashi M., Enoki T., Takeoka Y., Ishibashi T., Kuroda K., Tanaka K., Wang G., Grosberg A. Y., Masamune S., Tanaka T. Reversible adsorption of calcium ions by imprinted temperature sensitive gels. J. Chem. Phys. 2001;114 2812–2816.

[80] D'Oleo R., Alvarez-Lorenzo C., Sun G. A new approach to design imprinted polymer gels without using a template. Macromolecules 2001;34 4965–4971.

[81] Moritani T., Alvarez-Lorenzo C. Conformational imprinting effect on stimuli-sensitive gels made with an "imprinter" monomer. Macromolecules 2001;34 7796–7803.

[82] Ando N. Salt Effects on Adsorption affinities of random and imprinted copolymer gels. Senior Thesis. Massachusetts Institute of Technology, 2001.

[83] Hiratani H., Alvarez-Lorenzo C., Chuang J., Guney O., Grosberg A. Yu., Tanaka T. Effect of reversible cross-linker, N,N′bis(acryloyl)cystamine, on calcium ion adsorption by imprinted gels. Langmuir 2001;17 4431–4436.

[84] Stancil K. A., Feld M. S., Kardar M. Correlation and cross-linking effects in imprinting sites for divalent adsorption in gels. http://xxx.lanl.gov/abs/cond-mat/0301026 2003.

[85] Langmuir I. The adsorption of gases on plane surfaces of glass, mica and platinum. J. Am. Chem. Soc. 1918;40(9) 1361–1403.

[86] Atkins P. W. Physical Chemistry. Oxford: Oxford University Press, 1978.

[87] Kittel C., Kroemer H. Thermal Physics 2nd ed. New York: W. H. Freeman and Co., 1980.

[88] Bromberg L., Grosberg A. Yu., Matsuo E. S., Suzuki Y., Tanaka T. Dependency of swelling on the length of subchain in poly($N,N$-dimethylacrylamide)-based gels. J. Chem. Phys. 1997;106 2906–2910.

[89] Schild H. G. Poly($N$-isopropylacrylamide): experiment, theory and application. Prog. Polym. Sci. 1992;17 163–249.

[90] Hsein T. Y., Rorrer G. L. Heterogeneous cross-linking of chitosan gel beads: kinetics, modeling, and influence on cadmium ion adsorption capacity . Ind. Eng. Chem. Res. 1997;36 3631–3638.

[91] Eichenbaum G. M., Kiser P. F., Shah D., Meuer W. P., Needham D., Simon S. A. Alkali earth metal binding properties of ionic microgels. Macromolecules 2000;33 4087–4093.

[92] Grosberg A. Yu., Khokhlov A. R. Statistical Physics of Macromolecules. New York: AIP, 1994.

[93] Nikolov S., Doghri I. A micro/macro constitutive model for the small-deformation behavior of polyethylene. Polymer 2000;41(5), 1883–1891.

[94] Singhal A., Beaucage G., Harris M. T., Toth L. M., Keefer K. D., Lin J. S., Hu M. Z. C., Peterson J. R. Structure and growth kinetics of zirconium hydrous polymers in organic solutions. J. Non-cryst. Solids 1999;246 197–208.

[95]  Maeda M., Bartsch R. A. Molecular and ionic recognition with imprinted polymers. In Ref. [66] p.1–8.